\documentclass[usegraphicx, usenatbib,useAMS]{mn2e}
\usepackage[utf8]{inputenc}
\usepackage{epsfig}
\newcommand{\beq}{\begin{equation}}
\newcommand{\eeq}{\end{equation}}

\def\gtsima{$\; \buildrel > \over \sim \;$}
\def\ltsima{$\; \buildrel < \over \sim \;$}
\def\prosima{$\; \buildrel \propto \over \sim \;$}
\def\gsim{\lower.7ex\hbox{\gtsima}}
\def\lsim{\lower.7ex\hbox{\ltsima}}
\def\simgt{\lower.7ex\hbox{\gtsima}}
\def\simlt{\lower.7ex\hbox{\ltsima}}
\def\simpr{\lower.7ex\hbox{\prosima}}
\def\la{\lsim}
\def\ga{\gsim}

\newdimen\hssize
\newdimen\hdsize
\usepackage{amssymb}
\usepackage{amsmath}
\usepackage{longtable}
\usepackage{lscape}
\usepackage{color}
\usepackage{afterpage}
\usepackage{booktabs,fixltx2e}
\usepackage[flushleft]{threeparttable}

\begin{document}
\setlength{\hbadness}{10000}

\title[HII complexes and the PAH life cycle]
{Optical and infrared emission of H~{\sc ii} complexes as a clue to the PAH life cycle}

\author[M. S. Khramtsova et al.]
{M. S. Khramtsova$^{1}$\thanks{E-mail: khramtsova@inasan.ru},
D. S. Wiebe$^{1}$, T. A. Lozinskaya$^{2}$, O. V. Egorov$^{2}$ \\
$^{1}$ Institute of Astronomy, Russian Academy of Sciences, Pyatnitskaya str. 48, Moscow 119017, 
Russia\\
$^{2}$ Lomonosov Moscow State University, Sternberg Astronomical Institute, 13 Universitetskij prospekt, Moscow 119234, Russia}

\maketitle

\label{firstpage}

\begin{abstract}
We present an analysis of optical spectroscopy and infrared aperture photometry of more than 100 H~{\sc ii} complexes in nine galaxies. Spectra obtained with the 6-m telescope of SAO RAS are used along with archival data from {\em Spitzer\/} and several ground-based telescopes to infer a strength of polycyclic aromatic hydrocarbon (PAH) emission, age, properties of the UV radiation field, and metallicity of studied H~{\sc ii} complexes. Physical properties (age, radiation field parameters, metallicity) are related to the $F_{8}/F_{24}$ ratio used as a proxy for the PAH abundance in order to reveal factors that may influence the PAH evolution in H~{\sc ii} complexes. The well-known correlation between the $F_{8}/F_{24}$ ratio and metallicity is confirmed in the studied complexes. The infrared flux ratio also correlates with the [O~{\sc iii}]$\lambda 5007/\mathrm{H\beta}$ ratio which is often considered as an indicator of the radiation field hardness, but this correlation seems to be a mere reflection of a correlation between [O~{\sc iii}]$\lambda 5007/\mathrm{H\beta}$ and metallicity. In separate metallicity bins, the $F_{8}/F_{24}$ ratio is found to correlate with an age of an H~{\sc ii} complex, which is estimated from the equivalent width of $\mathrm{H}\beta$ line. The correlation is positive for low metallicity complexes and negative for high metallicity complexes. Analysing various mechanisms of PAH formation and destruction in the context of found correlations, we suggest that PAH abundance is likely altered by the UV radiation within H~{\sc ii} complexes, but this is not necessarily due to their destruction. If PAHs can also form in H~{\sc ii} complexes due to some processes like aromatisation, photodestruction, shattering and sputtering of very small grains, the net $F_{8}/F_{24}$ ratio is determined  by a balance between all these processes that can be different at different metallicities.
\end{abstract}


\section{Introduction}
The tiniest dust particles, polycyclic aromatic hydrocarbons (PAHs), play a crucial role in the evolution of the interstellar medium (ISM). They are an important ingredient in astrochemical reactions, in particular, catalysing H$_2$ formation \citep{LePage} and enriching the ISM with simple organic species, being destroyed in photodissociation regions (PDRs) \citep{Pety}. They absorb UV-emission and heat gas via photoelectric effect. However, the genesis of these particles is still far from being understood. It is generally believed that PAHs are synthesised in hot dense outflows ($\sim1000$\,K) of carbon-rich AGB stars \citep{GailSedlmayr87,Latter91,Cherchneff92,DhanoaRawlings} or in novae \citep{BodeEvans}. While theoretical calculations apparently confirm this possibility, firm observational evidence is still scarce. PAHs are not seen in the vicinity of AGB stars, except for a few cases when their emission is observed in AGB stars with hot companions \citep{Boersma}. The weak PAH emission of AGB stars without companions is probably explained by insufficient UV irradiation that is needed to excite PAHs. However PAHs are often observed in planetary nebulae \citep{pne2,pne1} that represent a later stage of AGB star evolution. On the other hand, it was proposed in a number of works \citep{LD02,Mattioda} that some PAHs can be excited even by UV-poor radiation field. If this is the case, the lack of PAHs emission close to AGB stars becomes an issue in the origin of these particles.

Even though some PAHs along with other dust types can be synthesized in AGB stars, this process may not be sufficient to produce all the dust, especially small grains, observed in our Galaxy and in other galaxies. For example, \citet{Matsuura2009} and \citet{Matsuura2013} showed that enrichment of the ISM by AGB stars in the LMC and SMC fails to explain the total dust and PAH abundance in these galaxies. PAH destruction in the ISM is apparently so effective \citep{mic2,mic1,mic3} that they do not survive the travel from parent stars to H~{\sc ii} complexes where they are observed. The misbalance between the dust formation and destruction has also been noted for our Galaxy by \citet{Jonesetal1994} and \citet{Jonesetal1996} \citep[see however][]{JonesNuth}. A conventional explanation is that some dust formation in molecular clouds is possible. A few mechanisms were suggested to explain PAHs and/or PAH precursors formation in cold molecular clouds \citep{Greenberg,Parker11}. Some PAHs can also make it into interiors of H~{\sc ii} complexes after having been formed in protoplanetary disks \citep{Woods,everett}.

A reverse process is also possible: PAHs may be a product of destruction of larger particles. \cite{Asano2013} and \cite{Hirashita2013} considered shattering process in the ISM as the main route to formation of small grains in galaxies. Besides shattering, other destruction mechanisms like sputtering, can be effective for larger particles in some circumstances. While gas sputtering works in hot plasma of supernova remnants at temperatures $>10^5$\,K \citep{DraineSalpeter}, alternative processes like chemical sputtering can be at work at lower temperatures \citep{Salonen}. \cite{jones13} has shown that amorphous carbonaceous grains can become aromatised under the influence of UV radiation. Finally, in H~{\sc ii} complexes and in PDRs formation of PAHs through UV evaporation of larger carbonaceous grains can be significant \citep{Berne,Pilleri}.

It is now well established that abundance of PAHs correlates with the metallicity of the environment where the PAH emission is observed \citep[e.g.][]{Draine2007,Smith2007,Wu2007,Hunt2010}. Two typical explanations relate this correlation either to metallicity-dependent formation \citep{Galliano2008} or to metallicity-dependent destruction \citep{Madden2000}. It has been shown in a number of studies that the correlation between the PAH abundance and metallicity holds not only globally, for a galaxy as a whole, but also locally, for individual H~{\sc ii} complexes within the same galaxy \citep{Gordon2008,ic10,Khramtsova}. This apparently indicates that the mechanism, responsible for the correlation, operates {\em in situ\/}, in the same locations where PAHs are observed. In this case it seems reasonable to expect a certain connection between the PAH abundance and the region age. \cite{Slater} studied PAH emission from classical H~{\sc ii} regions and superbubbles in the Large Magellanic Cloud and found no correlation between the emission properties and the morphology of an H~{\sc ii} complex despite significant differences in ionization conditions. They concluded that whatever dust processing occurs in H~{\sc ii} complexes it operates on a timescale that is either much shorter or much longer than the H~{\sc ii} complex evolutionary timescale.

In the work of \cite{Khramtsova} we performed an aperture photometry of about 200 star-forming complexes in 24 galaxies to find correlations between PAH emission properties. A local character of the PAH-metallicity correlation and an apparent anti-correlation between the PAH abundance and radiation field led us to conclude that the `destructive' scenario is a preferred one. However, to put any conclusion on the PAH evolution in H~{\sc ii} complexes on a more solid ground, a comparison with ages would be helpful.

In this work we attempt to relate the PAH content in extragalactic H~{\sc ii} complexes to their metallicity, shape of the ionizing UV field spectrum, and age. The correlations or a lack of thereof between the $F_{8}/F_{24}$ ratio and physical parameters of H~{\sc ii} complexes like an age and ionizing UV field strength may be a key to understand the formation and destruction of the PAHs. Specifically, if we give a preference to a `destructive' scenario, we might expect that the PAH content decreases with age in H~{\sc ii} complexes, and if PAHs are destroyed more effectively by harder radiation, we might also expect that the PAH content decreases faster when radiation field hardness is greater.

The structure of the paper is the following. In Section~2 we present observational data that are used in our analysis. In Section~3 we describe how physical parameters of H~{\sc ii} complexes are estimated from these data and also comment on the reliability of our estimates. In Section~4 we provide correlations between estimated parameters of H~{\sc ii} complexes and discuss their relation to the problem of the PAH evolution in Section 5.

\section{Data}

In this Section we present observational data on the optical spectroscopy of extragalactic H~{\sc ii} complexes used to derive their metallicities, ages, and radiation field properties and infrared data used to perform the aperture photometry of the same complexes.

\subsection{Sample}

We present results based on investigations of H~{\sc ii} complexes in nine galaxies. The sample was chosen to probe a correlation of $F_{8}/F_{24}$ ratio with parameters of H~{\sc ii} complexes in a wide range of metallicities. We obtained own observations for two galaxies, NGC~7741 and IC~1727, that represent low and high ends of the galactic metallicity distribution. Their spectroscopic observations are presented in this work for the first time. To make the sample representative we expanded it with seven more galaxies. These galaxies have different morphological types, metallicities and are included in the SINGS survey. Also, they were studied in our previous work \citep{Khramtsova}. There are other SINGS galaxies with spectroscopic data available, however, they all have quite high metal content. We have tried to make the sample as uniform as possible and not to bias the sample towards high metallicity galaxies. Spectroscopic data for H~{\sc ii} complexes in the added seven galaxies were taken from the literature. In total, we found spectroscopic data for 199 complexes in considered galaxies. Due to a lower resolution  and insufficient sensitivity of IR images not all complexes studied spectroscopically can be measured photometrically on IR images. We include in the final sample only those H~{\sc ii} complexes that both have spectroscopic observations available and can be detected and measured with sufficient signal-to-noise ratio (S/N $>3$ at 8\,$\mu$m) on IR images. Ninety complexes, mostly of small angular size, which have been observed spectroscopically, but cannot be resolved or are too weak on IR images, were excluded from the sample. They have nearly the same age and metallicity distributions as the remaining complexes and, thus, their exclusion does not introduce any strong bias into the sample.

Altogether the current sample includes 109 H~{\sc ii} complexes (including newly observed complexes in IC~1727 and NGC~7741). The information about the number of  H~{\sc ii} complexes studied in each galaxy along with the distance to the galaxy, its morphological type and a source of optical spectroscopic data (with corresponding references) is presented in Table~\ref{tabular:datasample}. The number of complexes in IC~1727 and NGC~7741 that have been studied only spectroscopically (see above) is shown in parentheses. 

\begin{figure}
    \includegraphics[width=0.45\textwidth]{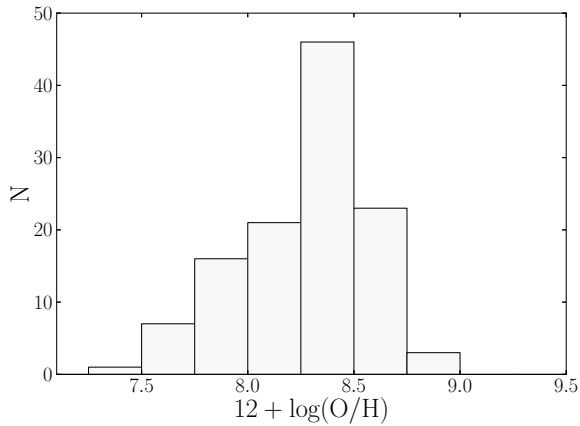}
    \caption{A metallicity distribution of studied H~{\sc ii} complexes. Metallicities were estimated by the `ONS'-method \protect{\citep{ONS}}.}
    \label{z_distr}
\end{figure}

The entire sample covers a wide range of metallicities from the lowest values of the order of 7.5--7.9 (complexes in Holmberg II) to solar or super-solar values, around 8.6--8.8 (e.g. NGC~3184). It allows studying how physical properties change over the transition from low to high metallicity environments. A metallicity distribution for our sample is shown in Fig.~\ref{z_distr}.

The nearest galaxy in the sample is Holmberg~II that belongs to the M81 group at a distance of about 3.4 Mpc \citep{Jacobs}. The most remote is NGC~7741, with distance estimates ranging from 12.1 to 17.5~Mpc \citep{Willick} and the most recent measurements being 15.1 Mpc \citep{Tully}. Therefore a typical aperture size of $6\arcsec$ for NGC~7741 corresponds to the linear size of about 500 pc, while a $10\arcsec$ aperture for the closest galaxy of the sample, Holmberg II, corresponds to $\sim200$ pc.

\begin{table*}
\caption{Summary of galaxies included in the work: name, number of studied H~{\sc ii} complexes, adopted distance and its reference source, morphological type and reference sources for optical spectroscopic data.}
\label{tabular:datasample}
\begin{center}
\begin{tabular}{ccccc}
\hline
Name    & Number of & Distance, & Morphological \\
        &H~{\sc ii} complexes& Mpc [ref.]    & type  & ref.\\ \hline
IC 1727    & 6 (11)     & 6.8 [1]   & SB(s)m & this work\\
NGC 7741   & 6 (10)     & 15.1 [1]     & SBc  & this work \\
Holmberg II& 12        & 3.4 [3]      & Im     & [6]\\
IC 2574    & 11        & 3.8 [4]      & SAB(s)m & [7]\\
NGC 628    & 10        & 7.3 [2]      & Sc & [8,9]\\
NGC 925    & 18        & 9.3 [2]      & Scd & [9]\\
NGC 3184   & 11        & 11.1 [5]     & SABc & [9]\\
NGC 3621   & 21        & 6.7 [2]      & SBcd & [10]\\
NGC 6946   & 14        & 5.9 [2]      & SABc & [11]\\
\hline
\end{tabular}
\end{center}

\begin{tablenotes}
 \small
      \item References: (1) \cite{Tully}, (2) \cite{Karachentsev}, (3) \cite{Jacobs}, (4) \cite{Dalcanton}, (5) \cite{leonard}, (6) \cite{Egorov12}, (7) \cite{Croxall}, (8) \cite{Gusev628}, (9) \cite{vanzee}, (10) \cite{Ryder}, (11) \cite{Gusev6946}
\end{tablenotes}
\end{table*}

\subsection{Spectroscopic Data}

\begin{figure*}
    \includegraphics[width=0.45\hdsize]{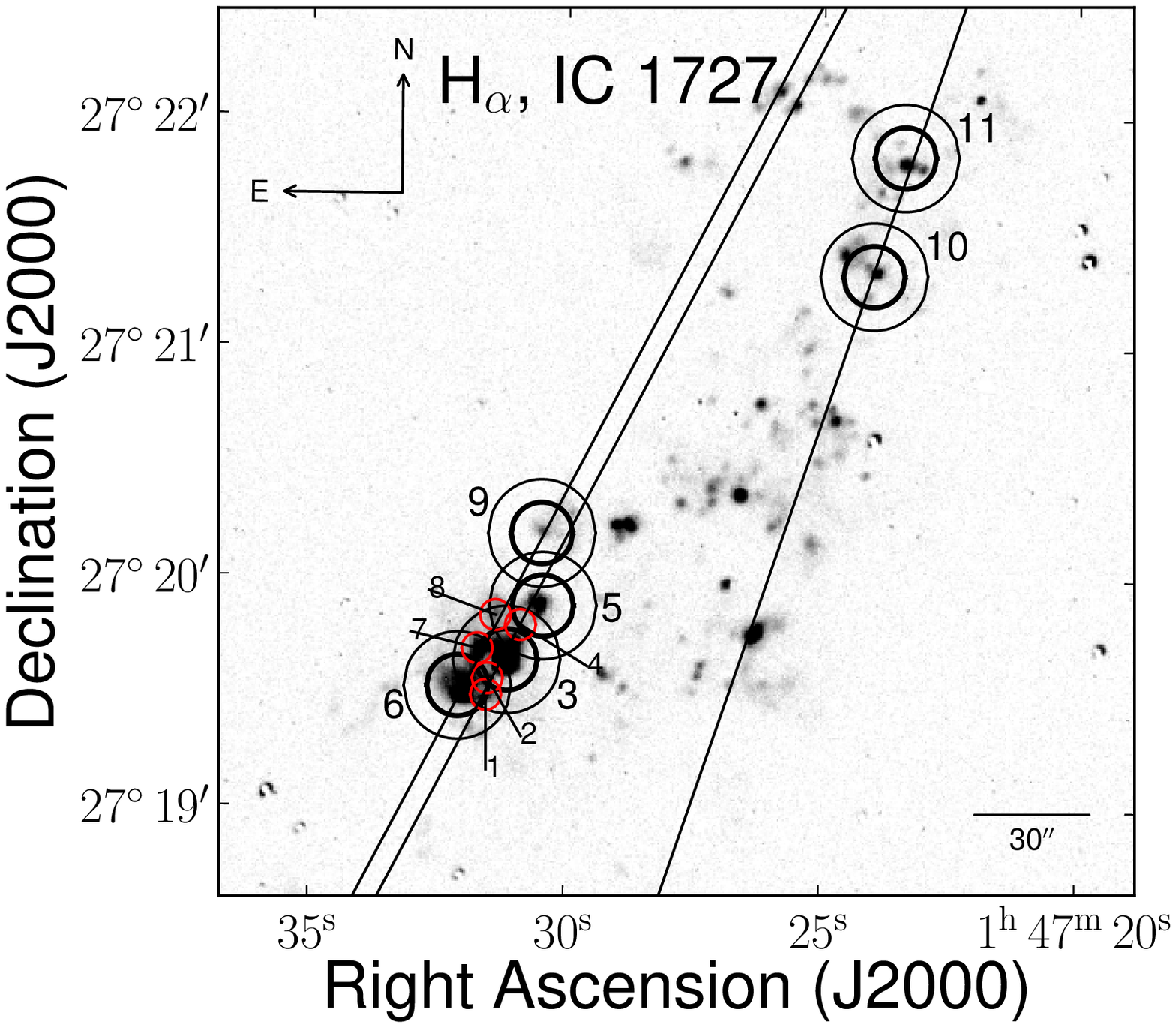}
    \includegraphics[width=0.45\hdsize]{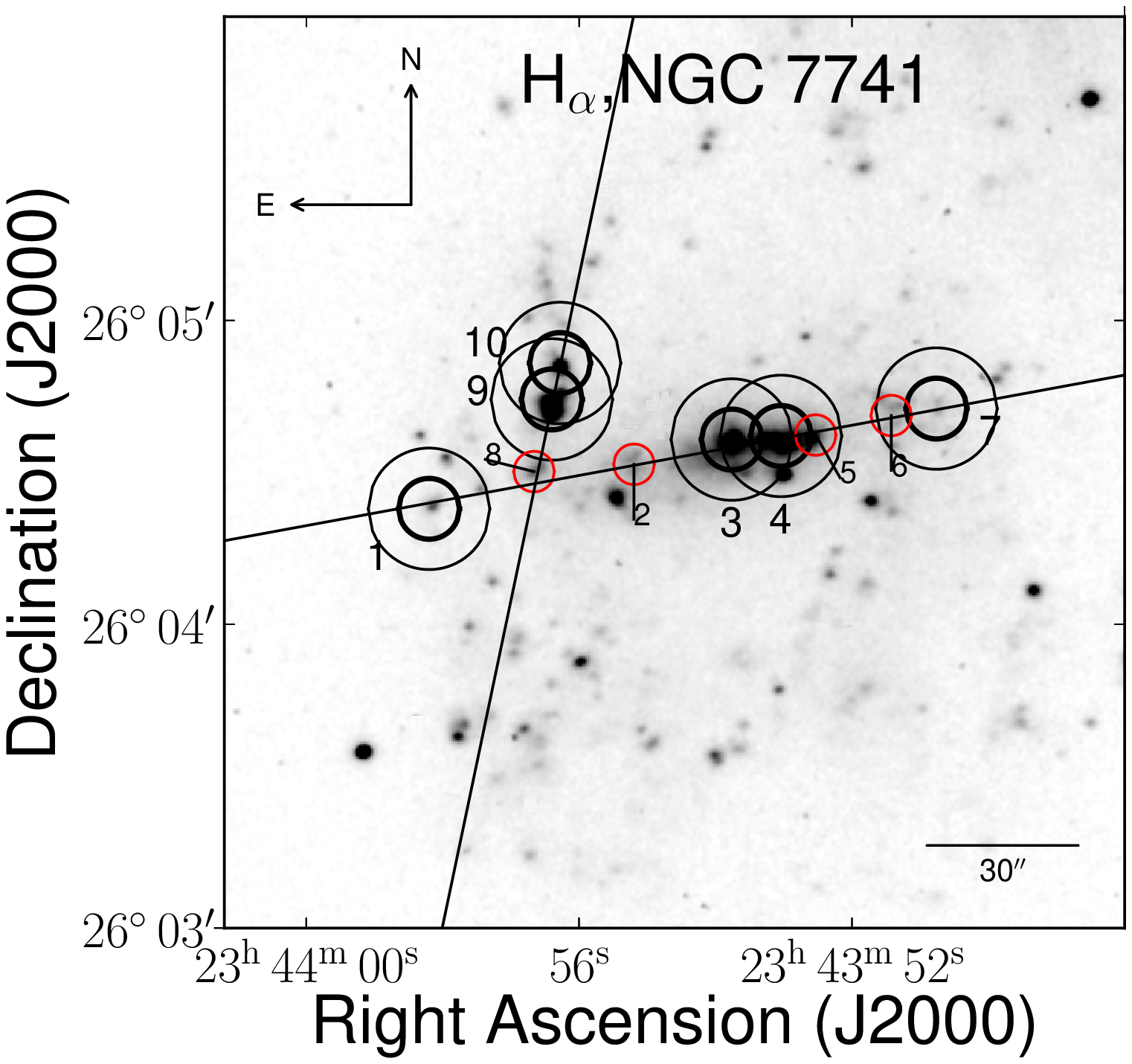}
    \caption{H$\alpha$ images of IC~1727 (left) and NGC~7741 (right) galaxies with locations of slits and H~{\sc ii} complexes. Red circles represent H~{\sc ii} complexes studied only spectroscopically. Black circles indicate complexes studied both spectroscopically and photometrically. Thick black circles indicate areas for performing photometry, while thin black circles show regions used for background subtraction.}
    \label{examples}
\end{figure*}

Spectroscopic observations of H~{\sc ii} complexes in IC~1727 and NGC~7741 galaxies were performed with the 6-m telescope of the Special Astrophysical Observatory of the Russian Academy of Sciences (SAO RAS) using the multi-mode SCORPIO focal reducer \citep{Afanasiev05} in the long-slit spectroscopy mode on 22--23 of October, 2011.  The used slit has length of 6$\arcmin$ and width of 1$\arcsec$. In Fig.~\ref{examples} we show images of IC~1727 and NGC~7741 galaxies in H$\alpha$ passband. Locations of slits and studied H~{\sc ii} complexes are also indicated. The complexes studied both spectroscopically and photometrically are marked by black circles, while red circles correspond to complexes studied only spectroscopically. We used VPHG550G grism and EEV CCD42-40 CCD detector with the 3700-7300\AA\ wavelength range that covers all the important nebular emission lines. The spectral resolution is 10\AA\ and dispersion is 2.1\AA/pix, which is sufficient for our goal of abundance and age estimating. Exposure time, 15~min, was the same for all spectra. The number of exposures for one slit location is 4. Seeing during observations varied from 0.8$\arcsec$ to 1.1$\arcsec$.

\begin{figure}
    \includegraphics[width=0.45\hdsize]{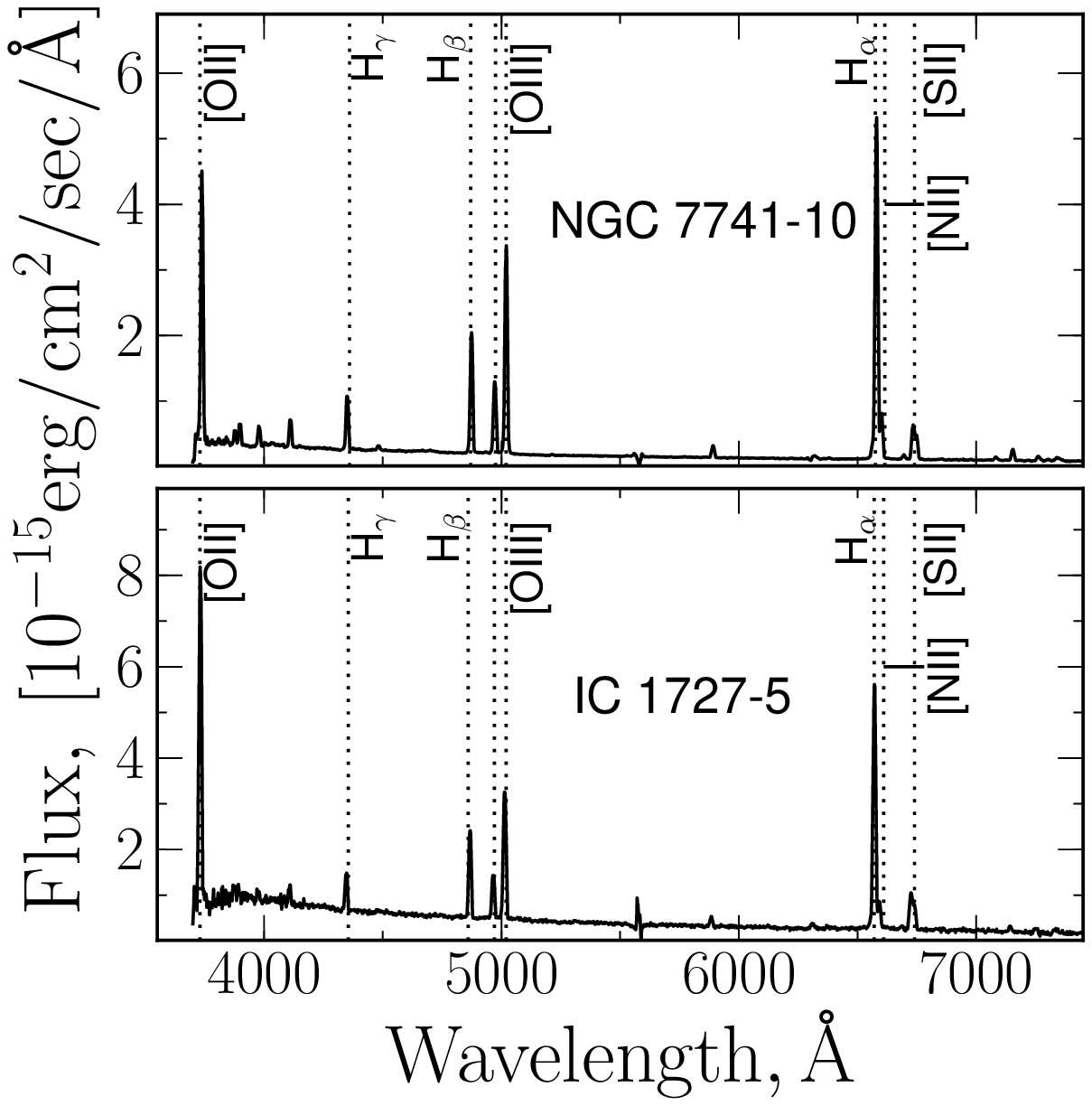}
    \caption{Typical spectra of H~{\sc ii} complexes in NGC~7741 and IC~1727 galaxies obtained with the 6-m telescope of the SAO RAS. Bright emission lines used for further analysis are marked on the spectra.}
    \label{examplessp}
\end{figure}

Spectra were reduced with a standard procedure that includes the bias field subtraction, normalization by flat fields, wavelength calibration, sky lines subtraction, and flux calibration using spectrophotometric standards GD 93-48 and AGK+81d266. All steps were performed with the \textsc{idl} software developed at the SAO RAS specifically for long-slit data. Emission lines were fitted by a single or multiple gaussians. [N~{\sc ii}]$\lambda\lambda6548,6584$ and H$\alpha$ lines were measured using a triple gaussian. Spectra were corrected for the interstellar extinction using the average decrement from the ratios of Balmer lines ${\rm H}\alpha/{\rm H}\beta$ and ${\rm H}{\gamma}/{\rm H}\beta$ when it is possible or ${\rm H}\alpha/{\rm H}\beta$ only if ${\rm H}{\gamma}$ is too weak. Theoretical values for these ratios, 2.86 and 0.47, correspondingly, were taken from \citet{osterbrock} for $T_e \approx 10000$K. We calculated colour excess E(B--V) using the extinction curve from the work of \cite{Cardelli} and parameterization from the work of \cite{fitzpatrick}. Examples of typical spectra for H~{\sc ii} complexes in these galaxies are presented in Fig.~\ref{examplessp}. We measured equivalent widths with the stellar continuum, while to measure line intensities, we subtracted it using the open software ULYSS\footnote{http://ulyss.univ-lyon1.fr/}, though stellar contribution for studied complexes is almost negligible and does not affect results noticeably.

The list of H~{\sc ii} complexes and results of the spectroscopy for IC~1727 and NGC~7741 galaxies are presented in Table~\ref{table:lines}. The offsets are given relative to the centre of the galaxy (positive values are toward the East and the North). The equatorial coordinates of IC~1727 and NGC~7741 centres, adopted from \citet{Saintonge}, are $\alpha = 01^{h} 14^{m} 29.90^{s}$, $\delta = +27^\circ 19\arcmin 58.0\arcsec $ and $\alpha = 23^{\rm h} 43^{\rm m} 54.5^{\rm s}$, $\delta = +26^\circ 04\arcmin 31.0\arcsec$, respectively. Position angle PA $= 150^\circ$, inclination $i = 68^\circ$, and radius at 25 mag in the B band $R_{25} = 193\arcsec$ for IC~1727 are taken from \citet{Epinat}. Parameters for NGC~7741 are taken from the HyperLeda galactic database\footnote{http://leda.univ-lyon1.fr} \citep{Paturel}: $\mathrm{PA} = 153.5^\circ$, $i = 50.2^{\circ}$, and $R_{25} = 108.9\arcsec$. Table~\ref{table:lines} includes the offsets relative to centres of galaxies, bright line intensities, normalised to H$\beta$ intensity, colour excesses, metallicities, and radial distances of H~{\sc ii} complexes. Methods of metallicity estimation are described below.

Spectra of H~{\sc ii} complexes in Holmberg~II were also obtained with the 6-m telescope of the SAO RAS and presented in the work of \cite{Egorov12}. We used their spectra to measure an equivalent width of the H$\beta$ line and adopted their estimations of metallicity. Spectroscopy data (line intensities and $EW({\rm H}\beta)$) for IC~2574 are taken from \citet{Croxall}, data for NGC~628 are partly from \cite{Gusev628} and partly from \cite{vanzee}, data for NGC~6946 are taken from \cite{Gusev6946}, and NGC~925 and NGC~3184 data are taken from \cite{vanzee}. Finally, NGC~3621 data are from \cite{Ryder}. We calculate metallicities and ages of H~{\sc ii} complexes using these data.

\subsection{Infrared photometry}

We perform the aperture photometry for H~{\sc ii} complexes in all nine galaxies using the {\em Spitzer\/} data at wavelengths of 3.6, 4.5, 5.8 and 8.0~$\mu$m (the IRAC instrument) and 24~$\mu$m (the MIPS instrument). IC~1727 and NGC~7741 data were downloaded from the Spitzer Data Archive\footnote{http://sha.ipac.caltech.edu}. Other galaxies belong to the SINGS sample \citep{KenSINGS}, and we downloaded the reduced images from the SINGS project website\footnote{http://sings.stsci.edu}. Prior to performing the aperture photometry we convolved images to a MIPS resolution at 24~$\mu$m ($6\arcsec$) with the IDL convolution procedure and kernels provided by \cite{aniano}. Also, we rescaled images to make pixel sizes equal to 1.5$\arcsec$ for all images. 

We sum fluxes of pixels within chosen apertures subtracting an average background level in a surrounding ring adjacent to an aperture (width of a ring is $\sim 6\arcsec$). Note that we do not apply the IRAC extended source correction factors considering H~{\sc ii} complexes as point sources. Since we convolve images to matched point spread functions (PSF), we do not need to use IRAC or MIPS aperture corrections. Examples of apertures (thick circles) and background rings (thin circles) are demonstrated in Fig.~\ref{examples} for complexes in IC~1727 and NGC~7741. Contributions of bright pixels (three times brighter than the standard deviation in a ring, $\sigma_{\rm ring}$) in a background area are ignored to avoid overestimating. These pixels can be related to neighbouring complexes. The total error of an aperture flux consists of the background noise level ($\sigma_{\rm ring}$) and an error due to the shift of an aperture centre $\sigma_{\rm pos}$ that can be comparable to the first error. We do not take into account a beam size correction in measurements. The same method was used in our previous work \citep{Khramtsova}, and more details can be found there.

It must be noted that our technique to compute $\sigma_{\rm ring}$ likely leads to an underestimate of the noise, since our pixels are smaller than the beam, and adjacent pixels are likely correlated. This does not affect our conclusions, as $\sigma_{\rm ring}$ is not used in the computations directly, serving only as a tool to avoid pixels belonging to neighbouring complexes. But still, uncertainties in Table~\ref{table:photometry} can actually be somewhat larger because of this.

It is believed that 8~$\mu$m emission arises mostly from vibrational modes of PAHs, and 24~$\mu$m emission mostly represents thermal emission of hot very small grains (VSGs) \cite[e.g.][]{DL07}. However, the exact contributions of PAHs to the 8~$\mu$m emission and VSGs to the 24~$\mu$m emission may vary from an object to an object depending on a number of factors (e.g. on metallicity). Actually, thermal emission can contribute at $8\,\mu$m as well \citep{Engelbracht05}, and PAH emission can constitute a substantial fraction of emission at $24\,\mu$m \citep{Robitaille}. Stars and dust continuum from larger dust grains can contribute at these wavelengths as well. We estimated aromatic feature emission (`afe') at 8~$\mu$m and flux with subtracted stellar continuum (`ns') at 24~$\mu$m as suggested in the work of \cite{marble}. Note that throughout the paper we use shorter designation $F_{8}/F_{24}$ that actually means $F_{8}^{\rm afe}/F_{24}^{\rm ns}$.

Photometry results are presented in Table~\ref{table:photometry} that includes a designation of a complex (column 1), equatorial coordinates on the Spitzer images (columns 2,3), aperture radius (column 4), measured fluxes at 3.6, 4.5, 5.8, 8.0~$\mu$m, contribution from aromatic features at 8.0~$\mu$m, measured flux at 24~$\mu$m and flux without the stellar contribution at 24~$\mu$m (columns 5-11). Designations of complexes for IC~1727 and NGC~7741 are the same as in Table~\ref{table:lines}. Some of the complexes that we observed spectroscopically in NGC~7741 and IC~1727 are not resolved in Spitzer images, so no photometry for them is presented in Table~\ref{table:photometry}. Region designations for other galaxies were taken from the corresponding references with spectroscopic measurements.

\section{Physical parameters of H~{\sc ii} complexes}

The goal of this study is to check whether correlations exist between the H~{\sc ii} complex properties and their PAH content. In particular, it is interesting to see if PAH evolution is somehow related to the  radiation field properties in various environments. It has been proposed that harder radiation field in lower metallicity medium can be an underlying reason for the observed correlation between metallicity, 12~+~log(O/H), and the PAH mass fraction, $q_{\rm PAH}$. If PAHs are indeed destroyed faster by harder photons, at a given age we may expect less PAHs in an H~{\sc ii} complex with harder radiation field. Optical spectra presented in the previous Section allow estimating physical parameters of the studied H~{\sc ii} complexes, first of all, their metallicities and ages. Also, ratios between some spectral line intensities may depend on the hardness of ionizing field. Below we describe how we estimate these parameters, present our results that are summarised in Table~\ref{table: key results}, and also provide some comments on their reliability.

\subsection{Abundance of PAHs}

According to the dust emission model by \cite{DL07}, the PAH mass fraction, $q_{\rm PAH}$, in the total dust mass can be estimated, provided far-infrared fluxes are known for the studied region at least up to 160\,$\mu$m. This implies usage of {\em Herschel\/} data. However, most star-forming complexes we analyse are located in crowded environments so that {\em Herschel\/} angular resolution is not sufficient to study them separately from neighbouring regions. Thus, we limit ourselves with {\em Spitzer\/} data and use the $F_{8}/F_{24}$ ratio as a tracer of the PAH abundance. 

In our previous work we showed that $F_{8}/F_{24}$ ratio correlates with $q_{\rm PAH}$ \citep{Khramtsova}. Similar conclusions were reached by \citet{Sandstrom} and \citet{Engelbracht05} though the scatter can be significant, for example, because of differences in starlight fields.

The $F_{8}/F_{24}$ ratio is not a direct measure of $q_{\rm PAH}$ as it does not contain information on the cold dust component consisting of big grains (BGs). Nevertheless, we believe that the ratio can be used as a tracer of the PAH content for following reasons. According to \cite{DL07}, the fraction of 8\,$\mu$m emission in the total infrared flux sensitively depends on $q_{\rm PAH}$, while the relative intensity of the 24\,$\mu$m emission is only weakly dependent on the PAH content, and this dependence becomes even less pronounced for the radiation field parameter $U$ greater than ten ($U$ is the minimum starlight intensity in the complex in units of the interstellar radiation field in the Sun's vicinity), which is expected in star-forming regions. Apparently, the 24\,$\mu$m emission ratio to the emission of colder BGs is relatively constant for objects in our sample. Therefore $F_{8}/F_{24}$ ratio is equivalent to ratio of $F_{8}$ to the total infrared flux. Consequently, the ratio of 8\,$\mu$m and 24\,$\mu$m fluxes should reflect the PAH content reliably.

\subsection{Metallicities}

A number of methods have been developed over the years to estimate metallicity of H~{\sc ii} regions. It is believed that the most direct way to estimate metallicity is through the use of electron temperature $T_{e}$ \citep{Stasinska_meth}. To apply this method, an [O~{\sc iii}]$\lambda$4363 line is usually used. However, in most cases it is quite weak, so other methods that require only bright lines were proposed as alternatives to the `direct' one. They can be separated into empirical \citep[][etc.]{PT05,ONS,NS} and theoretical \citep[e.g., ][]{KK04} groups. A comparison of various methods and their relation to the `direct' method were studied extensively \citep{kewleym,Lopez,Egorov12}.

\begin{figure}
    \includegraphics[width=0.45\hdsize]{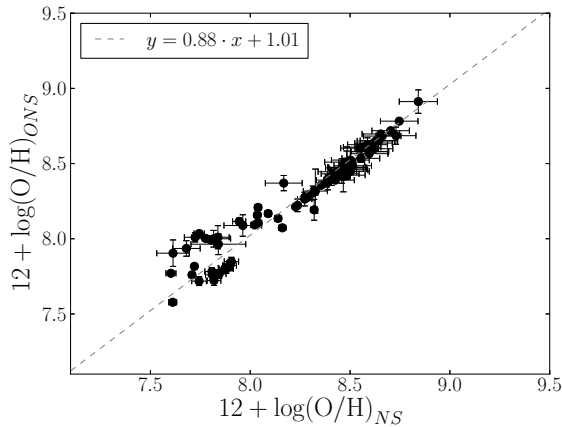}
    \caption{Comparison between the two metallicity estimation methods (NS and ONS) used in this work. The dotted line indicates the linear fit for this comparison. The corresponding equation is also shown.}
    \label{ns_vs_ons}
\end{figure}

We prefer to use a so-called `ONS' method based on the measurements of bright oxygen ([O~{\sc ii}]$\lambda\lambda 3727,3729$, [O~{\sc iii}]$\lambda\lambda 4959,5007$), nitrogen ([N~{\sc ii}]$\lambda\lambda6548,6584$), and sulfur ([S~{\sc ii}]$\lambda\lambda 6717,6731$) lines \citep{ONS}. In Holmberg~II data on the [O~{\sc ii}]$\lambda\lambda 3727,3729$ lines are not available except for a couple of complexes, so we rely on the `NS' calibration \citep{NS} to estimate metallicities in this galaxy. These two calibrations are close to each other as shown in the work of \cite{Egorov12}. We compared the NS and ONS metallicity estimates for the rest of our sample of H~{\sc ii} complexes and found them to be consistent with each other. This is illustrated in Fig.~\ref{ns_vs_ons}. The linear fit line is also shown.

\begin{figure*}
    \includegraphics[width=0.8\hdsize]{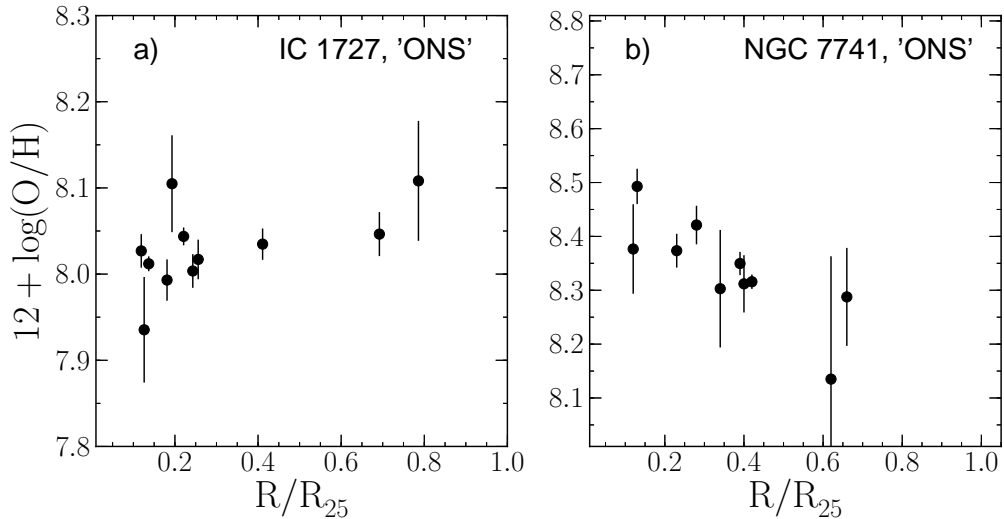}
    \caption{Metallicities vs. radial distances of H~{\sc ii} complexes for IC~1727 and NGC~7741. 
Metallicities are estimated by the ONS method.}
    \label{gradients}
\end{figure*}

Diagrams showing radial metallicity profiles in IC~1727 and NGC~7741 galaxies are presented in Fig.~\ref{gradients}. H~{\sc ii} complexes in disk galaxies typically demonstrate a metallicity gradient \citep{Vila,Zaritsky,vanzee}. This gradient is clearly seen in the NGC~7741 galaxy. A slope of a gradient likely depends on the galaxy morphological type and becomes steeper in late-type galaxies \citep{Oey,Zaritsky}. Unlike disk galaxies, SB(s)m galaxies are not expected to have a noticeable gradient, and indeed nearly flat oxygen abundance distribution is seen in IC~1727. Gradients for other galaxies in our sample have been presented in their respective studies except for Holmberg~II and IC~2574 where the gradient is not expected, in particular, because of rather compact location of studied H~{\sc ii} complexes.

\subsection{Ages}

We estimate ages of H~{\sc ii} complexes using $EW({\rm H}\beta)$. In a case of a single stellar population formed in an instantaneous star formation burst, $EW({\rm H}\beta)$ has a maximum at the onset of the star formation and then gradually decreases with time. This method was originally suggested by \cite{Copetti} and later discussed by \cite{Schaerer} and \cite{Stasinska_age}. \cite{Levesque10} gave some expressions to determine the burst age for different metallicities ($Z/Z_{\sun} = 0.05,0.2,0.4,1.0,2.0$). We use these expressions to estimate ages of H~{\sc ii} complexes with calibrations for an appropriate metallicity, even though an assumption of a single stellar population may not be strictly valid. If stellar populations of different ages coexist in the same H~{\sc ii} complex \citep{Bresolin99, Peacock13}, older stars can make a significant contribution to the continuum, correspondingly decreasing $EW({\rm H}\beta)$. This can lead to an age overestimation. In principle, multiple stellar populations can be traced by absorption lines and the increased slope of the spectra, but our spectra do not contain strong absorption lines, and the slopes are quite flat, so we use a single stellar population model to estimate ages of our H~{\sc ii} complexes. Potentially, this can be a problem for NGC~7741 as this galaxy is located further away than other galaxies from our sample so that it is possible that some older stars that do not belong to H~{\sc ii} complexes have been covered by the slit. In this case ages of these complexes are somewhat overestimated.

One possible source of uncertainty in our study is the potential inconsistency between metallicities that are used to derive the age calibration and metallicities estimated from spectral data. To estimate the role of this uncertainty, we utilize models presented by \cite{2010AJ....139..712L}\footnote{Available at http://www.emlevesque.com}. These authors have used the Starburst99 code \citep{sb99} to compute synthetic ionizing FUV radiation spectra as a function of metallicity, star formation history, and age. These spectra have then been used as an input for the Mappings III photoionization code \citep{mappings1,mappings2} to compute (in particular) line intensity ratios. 

We have used these line ratios to compute metallicities with NS and ONS calibrations. They turned out to be consistent with `true' metallicities given in the \cite{2010AJ....139..712L} grid as input parameters. Inconsistency only appears in models with the lowest value of the ionization parameter ($10^7$ cm s$^{-1}$; see the next subsection) and at ages greater than 5 Myr, where NS and ONS methods tend to give lower metallicities than the `true' ones.

The similar analysis has been presented in the work of \cite{Lopez}. These authors have also shown that in the case of continuous star formation NS and ONS methods provide consistent, but underestimated metallicities. In order to check the significance of this underestimation, we have repeated all our computations with metallicities artificially increased by 0.5 dex (a typical discrepancy in the \cite{Lopez} study) and found that none of our conclusions is affected by this change. So, to avoid confusion, we provide age values computed with the ONS (NS) metallicities.

Ages for some regions in IC~2574 and Holmberg~II were estimated earlier in the works of \cite{Stewart_ic} and \cite{Stewart_ho}, correspondingly. Their method is based on the assumption of a single stellar population as well, but measurements FUV, B, and H$\alpha$ are used instead of H$\beta$. We compare our age estimates with ages presented in cited works in Fig.~\ref{comparing}. In the work of \cite{Stewart_ho} all complexes are separated into four groups by their ages. We illustrate these estimates by squares with error bars that show age ranges for each group. Circles represents ages of complexes in IC~2574 from \cite{Stewart_ic}{\footnote{Note that in Table~2 from this paper the column with ages is actually a copy of the column with FUV luminosities. The authors provided us with the correct values of ages.}.

With the one exception, our results are consistent with previous estimates. The deviant point corresponds to an HSK61 complex in Holmberg~II. This complex is quite extended, so it is possible that a spectrograph slit was located in a position with older populations. Also, weakness of emission in this complex might have led to incorrect results. The kinematic age estimate for this region is only about 3.7 Myr \citep{wiebe14}. Otherwise, all age estimates are mainly in agreement.

\cite{pellerin} presented some age estimates for complexes in IC~2574 using colour-magnitude diagram for resolved stars. They divided all stellar systems into three age groups, with clusters from the youngest group having ages less than 10 Myr. All our complexes fall into this group. Apparently, they do not contain significant older populations, and, thus, an assumption of a single stellar population is justified for complexes in IC~2574. 

We have excluded the +085--042 region from NGC~3184 which has an exceptionally wide H$\beta$ line, according to \cite{vanzee}, corresponding to age of 0.1 Myr. We believe this age value is unreliable as in other aspects the region does not differ from the older regions.

\begin{figure}
\includegraphics[width=0.45\hdsize]{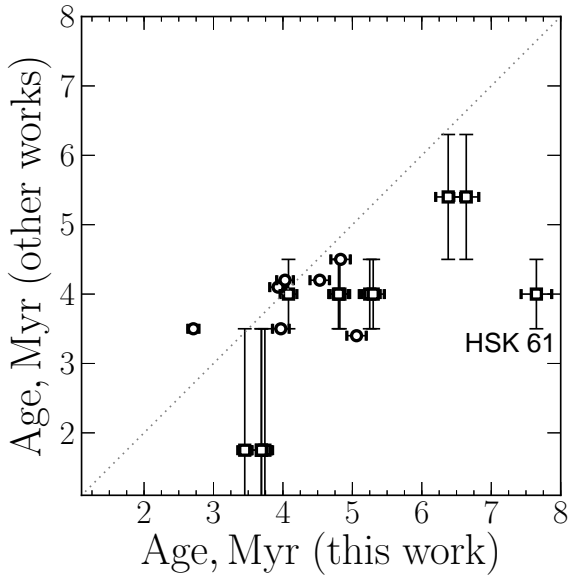}
\caption{Comparison of age estimates from this work and from the works of \protect\cite{Stewart_ic} and \protect\cite{Stewart_ho} for complexes in IC~2574 and Holmberg~II. Squares with error bars correspond to complexes in Holmberg~II. The real ages of complexes are contained within the ranges indicated by error bars and squares are just middle values for these ranges. Circles correspond to complexes in IC~2574.}
\label{comparing}
\end{figure}

\subsection{Radiation field hardness}

PAHs and other dust grains are destroyed by UV radiation, and it is reasonable to assume that the greater the average photon energy, the more effective it is as a destruction factor. Thus, the rate of PAH photodissociation should depend on the radiation field hardness. Some line ratios, like [O~{\sc iii}]$\lambda 5007/\mathrm{H\beta}$, [S~{\sc ii}]$\lambda 6717+6731/\mathrm{H\alpha}$, [Ne~{\sc iii}]/[Ne~{\sc ii}], are proposed as indicators of the shape of the radiation field spectrum. Line intensity ratios [O~{\sc iii}]$\lambda 5007/\mathrm{H\beta}$ and [S~{\sc ii}]/H$\alpha$ have been proposed as tracers of the radiation field hardness by \cite{2010AJ....139..712L}. Harder radiation increases O$^{++}$ abundance, as it contains more photons able to ionize O$^{+}$, and decreases S$^{+}$ abundance, as it penetrates deeper into the surrounding gas and produces more S$^{++}$ ions. Thus, [O~{\sc iii}]$\lambda 5007/\mathrm{H\beta}$ is expected to be an increasing function of the radiation field hardness, while [S~{\sc ii}]/H$\alpha$ should decrease in harder radiation field.

To verify if it is indeed the case, we use the pre-computed line ratios from the \cite{2010AJ....139..712L} grid and relate them to UV spectra properties. As the spectra themselves are not available in the published database, we have computed them using the Starburst99 code and the same parameter sets as in \cite{2010AJ....139..712L}. Specifically, instantaneous star formation burst is used with the Geneva ``Standard'' mass-loss tracks \citep{tracks1,tracks2,tracks3,tracks4}. Data for the electron density of $10^2$~cm$^{-3}$ are used.

Various definitions of the radiation field hardness (or alternatively softness) are described in the literature, like the starlight effective temperature \citep{vilchezpagel} or the spectral index in the range from 1 to 4 rydberg \citep[e.g., ][]{kewleyr}, often bound to the ionization state of certain elements. Our study is not directly related to the ionization structure of the complexes. Considering PAH destruction, we mostly need to know the overall energy distribution over hard and soft UV ranges. So, we define hardness as the ratio of integrated fluxes in the ranges from 90\AA\ to 912\AA\ and from 912\AA\ to 2000\AA. The intermediate value has no particular meaning and just separates the hard part of the spectrum that evolves significantly with time and the soft part of the spectrum longward of the Lyman limit that stays relatively constant. The adopted hardness measure, computed from the Starburst99 spectra, drops both with age and metallicity as shown in Fig.~\ref{hardness}. The similar definition was suggested in the work of \cite{lebouteiller14} for the PAH destruction investigation.

\begin{figure*}
\includegraphics[width=0.8\hdsize]{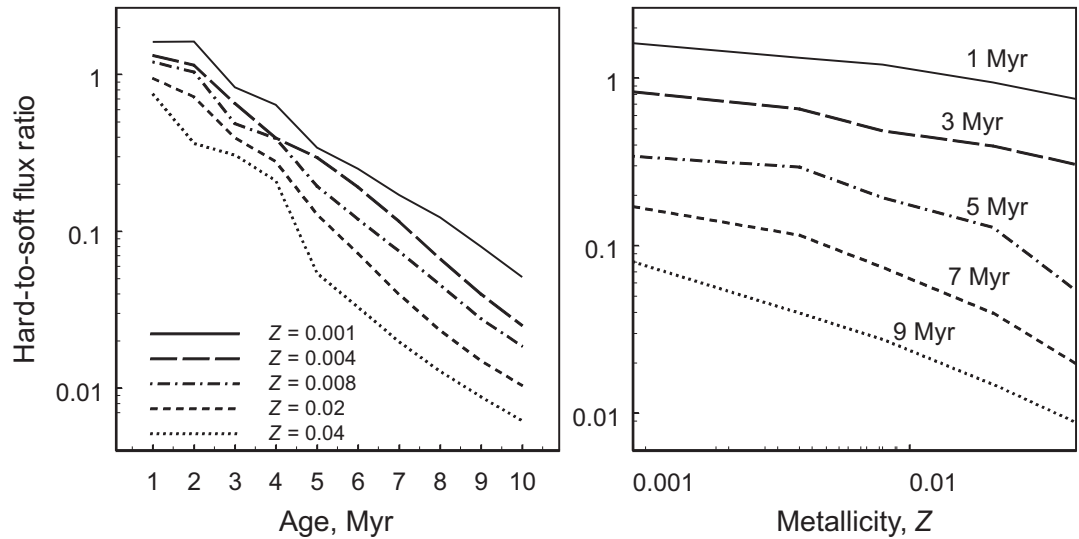}
\caption{Evolution (left panel) and dependence on metallicity (right panel) of the hard-to-soft flux ratio that we adopt as a measure of the radiation field hardness. Different linestyles correspond to different metallicities on the left panel and to different ages on the right panel. Spectra are computed with the Starburst99 code.}
\label{hardness}
\end{figure*}

The next step is to check whether the line ratios mentioned above can be used to trace the adopted hardness measure. An additional factor that enters the line ratio computation is the so-called ionization parameter $q$, which represents the ratio of the ionizing photon flux to the electron density \citep{osterbrock}. In the work of \cite{2010AJ....139..712L} seven values of $q \approx 10^7$, $2\times10^7$, $4\times10^7$, $8\times10^7$, $1\times10^8$, $2\times10^8$, and $4\times10^8$~cm s$^{-1}$ are used, which bracket possible ionization parameter values in actual H~{\sc ii} complexes.

\begin{figure}
\includegraphics[width=0.45\textwidth]{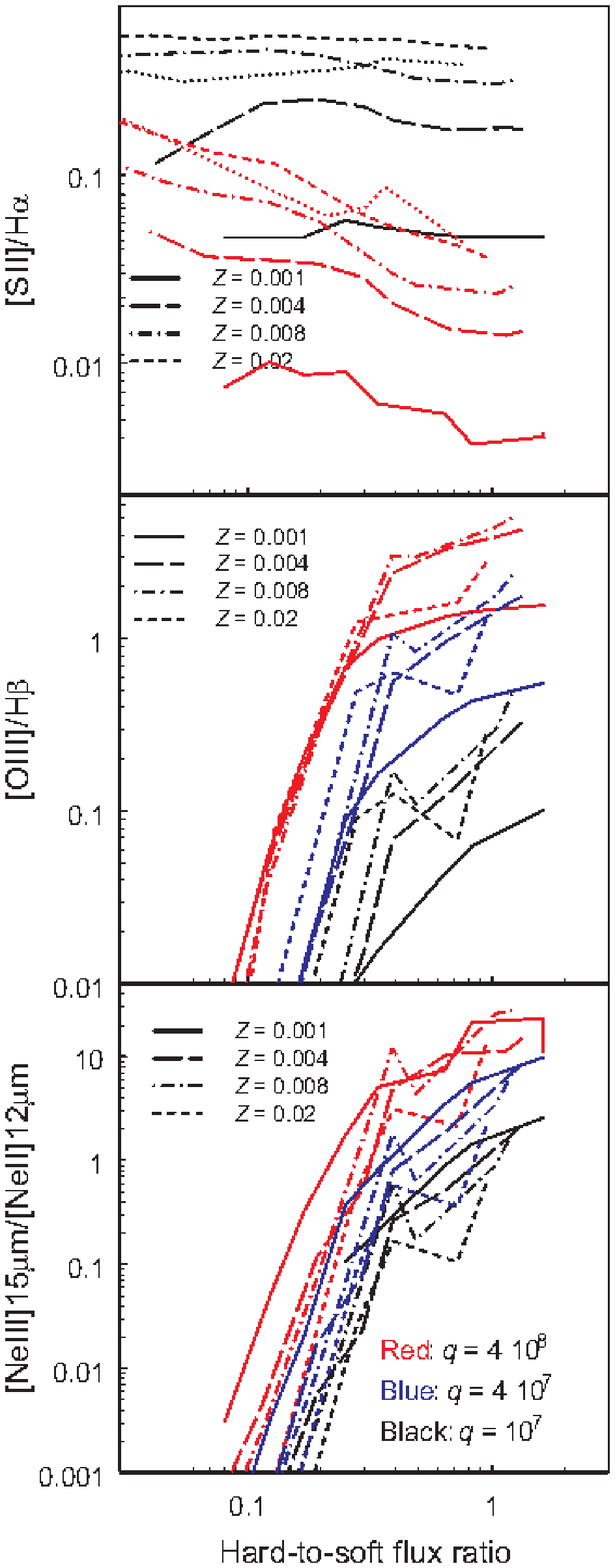}
\caption{The [S~{\sc ii}]/H$\alpha$, [O~{\sc iii}]$\lambda 5007/\mathrm{H\beta}$, and [Ne~{\sc iii}]/[Ne~{\sc ii}] line ratios in models with various values of the hard-to-soft flux ratio. Data are taken from the \protect\cite{2010AJ....139..712L} model grid. Different linestyles correspond to different metallicity values as indicated in the legends. Line colours on each panel indicate different value of the ionization parameter. Red lines correspond to $q=4\times10^8$~cm s$^{-1}$ (the maximum value in the Levesque et al. grid), blue lines correspond to $q=4\times10^7$~cm s$^{-1}$, black lines correspond to $q=10^7$~cm s$^{-1}$ (the minimum value in the Levesque et al. grid).}
\label{ratioOS}
\end{figure}

In Fig.~\ref{ratioOS} we show how the [S~{\sc ii}]/H$\alpha$, [O~{\sc iii}]$\lambda 5007/\mathrm{H\beta}$, and [Ne~{\sc iii}]/[Ne~{\sc ii}] ratios are related to our hard-to-soft flux ratio. Black and red curves correspond to $q$ values of $10^7$ and $4\times10^8$~cm s$^{-1}$, respectively. There is some anti-correlation between the spectrum hardness and the [S~{\sc ii}]/H$\alpha$ ratio for the higher value of $q$, as seen on the top panel, however, any correlation is absent for the lower value of the ionization parameter. The correlation between the overall radiation field shape and the [O~{\sc iii}]$\lambda 5007/\mathrm{H\beta}$ and [Ne~{\sc iii}]/[Ne~{\sc ii}] is much stronger, but apparently we need to estimate the ionization parameter in order to infer the radiation field hardness from the line ratios.

An ionization parameter in a particular H~{\sc ii} complex depends on many parameters, both internal and external \citep[e.g., ][]{dopita}. In principle, in our complexes it may take any value within the general limits indicated by observations of extragalactic H~{\sc ii} regions. \cite{lev14} have presented a parameterization of the log([O~{\sc iii}]$\lambda5007$/[O~{\sc ii}]$\lambda3727$) ratio as a diagnostic of the ionization parameter. Using their relation with the spectroscopic data we have collected and our metallicity estimates, we find that most our regions have $q$ below $4\times10^7$~cm~s$^{-1}$. Curves corresponding to this value of $q$ are shown with blue colour in Fig.~\ref{ratioOS}. Apparently, a ratio of neon infrared emission lines is the best tracer of the radiation field hardness, which was used in a number of works \citep[e.g.][]{lebouteiller14}. The scatter in the [O~{\sc iii}]$\lambda 5007/\mathrm{H\beta}$ line ratios for a given value of the hard-to-soft flux ratio is quite significant even in the narrow $q$ range from $10^7$~cm~s$^{-1}$ to $4\times10^7$~cm~s$^{-1}$. This implies that when the hard-to-soft flux ratio is high (at low metallicity and/or in younger complexes), the uncertainty in its value can reach an order of magnitude, which is comparable to scatter in observed line ratios.

To conclude this section, in Fig.~\ref{ohb} we relate the observed [O~{\sc iii}]$\lambda 5007/\mathrm{H\beta}$ ratio to our derived H~{\sc ii} complex metallicities and ages and compare it to the theoretical predictions. A top row shows that the line ratio is nearly independent of metallicity at 12~+~log(O/H)~$\la8.3$ and then decreases indicating that higher metallicity complexes mostly have softer radiation field. This is the same kind of behaviour that is predicted by theoretical models. Specifically, in panel Fig.~\ref{ohb}a we show results of numerical modelling by \cite{2010AJ....139..712L} for various ages and ionization parameters as indicated in the legend.

The correlation of the [O~{\sc iii}]$\lambda 5007/\mathrm{H\beta}$ ratio with age is weaker (Fig.~\ref{ohb}b). Note that there are few H~{\sc ii} complexes that possess high [O~{\sc iii}]$\lambda 5007/\mathrm{H\beta}$ ratios at ages of the order of 6~Myr or greater, contrary to theoretical predictions. These are, probably, the complexes where the instantaneous star formation burst assumption is not valid.

Fig.~\ref{ohb} also hints at uncertainty in $q$ estimates. Blue lines in Fig.~\ref{ohb}a and Fig.~\ref{ohb}b correspond to $q=4\times10^7$~cm~s$^{-1}$, which is higher than average in our complexes according to parameterization from the work of \cite{lev14}. However, models with this $q$ value predict lower [O~{\sc iii}]$\lambda 5007/\mathrm{H\beta}$ than in most considered complexes. In order to match theoretical predictions with observed values, $q\sim10^8$~cm~s$^{-1}$ or greater is needed as indicated by green lines in Fig.~\ref{ohb}a Fig.~\ref{ohb}b. Note that this discrepancy arises even though same observations and same model grids are used both in the [O~{\sc iii}]$\lambda 5007/$[O~{\sc ii}]$\lambda 3727$ parameterization and in Fig.~\ref{ohb}. \cite{lev14} used Geneva `high'-mass-loss stellar evolutionary tracks \citep{Meynetetal1994} to derive their parameterization, but in Fig.~\ref{ohb} higher $q$ values provide better match between theory and observations for both choices (`standard' or `high') of evolutionary tracks. If $q$ values of the order of $10^8$~cm~s$^{-1}$ or higher are indeed wide-spread, the [O~{\sc iii}]$\lambda 5007/\mathrm{H\beta}$ as an indicator of the radiation field hardness becomes more reliable.

\begin{figure*}
\includegraphics[width=0.9\hdsize]{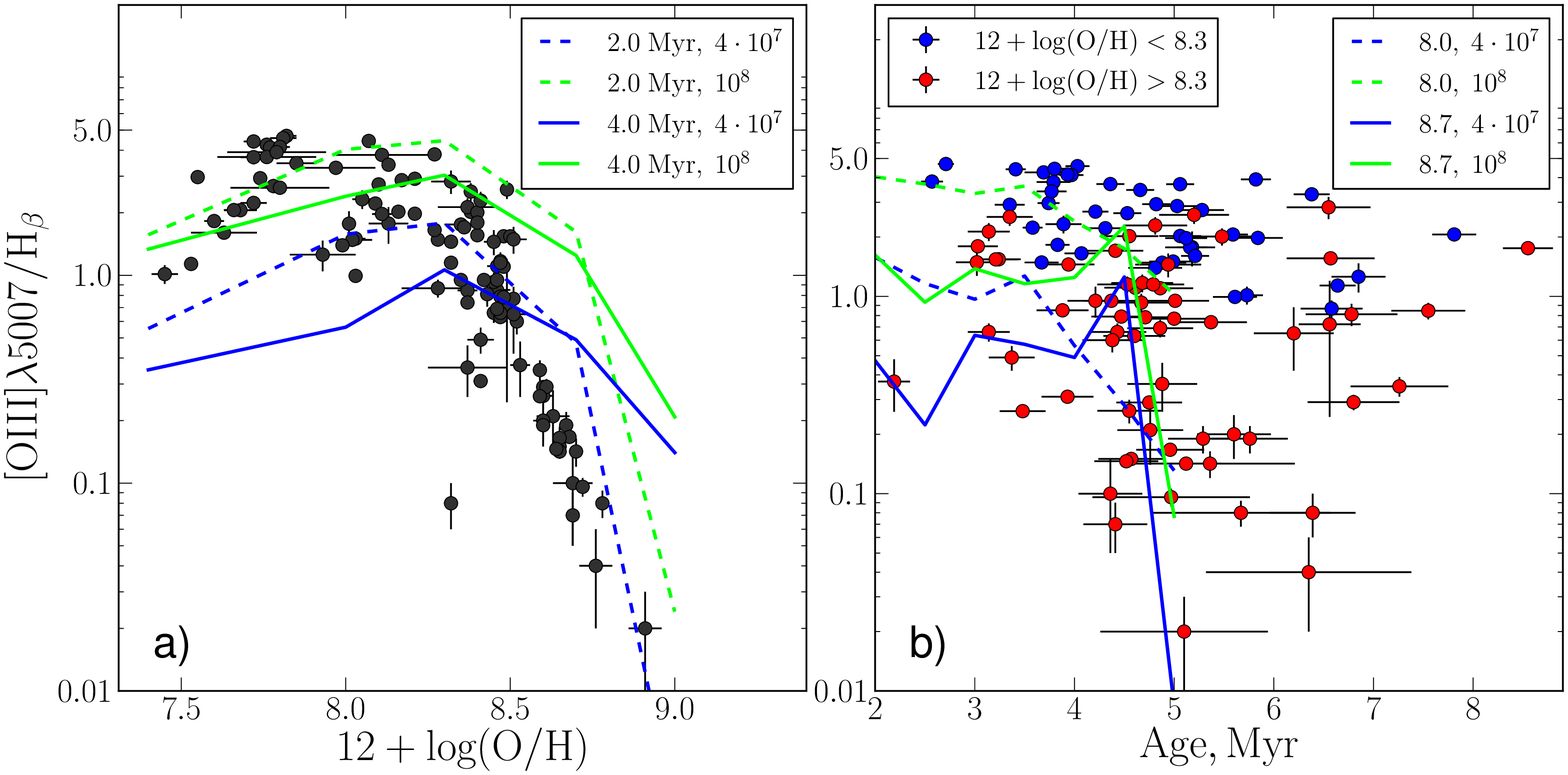}
\caption{a) Relation between the [O~{\sc iii}]$\lambda 5007/\mathrm{H\beta}$ ratio and the H~{\sc ii} complex metallicity. Solid and dashed lines show the relations computed with \protect\cite{2010AJ....139..712L} grids for ages of 2 and 4 Myr and ionization parameters of $4\cdot10^7$ (blue lines) and $10^8$~cm s$^{-1}$ (green lines). b) Relation between the [O~{\sc iii}]$\lambda 5007/\mathrm{H\beta}$ ratio and age. Regions with metallicity higher than 8.3 are marked by red circles while the rest regions are marked by blue circles. Solid and dashed lines again show relations modelled using \protect\cite{2010AJ....139..712L} grids for metallicities 8.0 (dashed lines) and 8.7 (solid lines) and ionization parameters of $4\cdot10^7$ (blue lines) and $10^8$~cm s$^{-1}$ (green lines).}
\label{ohb}
\end{figure*}

Summarising, all the considered theoretical line ratios ([S~{\sc ii}]/H$\alpha$, [O~{\sc iii}]$\lambda 5007/\mathrm{H\beta}$, and [Ne~{\sc iii}]/[Ne~{\sc ii}]) show expected dependence on the radiation field hardness, but none of them is particularly useful for our purpose due to the uncertainties related to the ionization parameter.

\section{Relations between various parameters}

Our `ideal' object is a star-forming complex that has been formed in a single star formation event and since then progressively evolves into an H~{\sc ii} complex. It is characterized by certain metallicity and by some initial PAH content that may or may not depend on metallicity. The oldest complex in our sample has age less than 10 Myr, so we do not expect to see any metallicity evolution. The initially abundant PAHs are destroyed by UV radiation, with harder photons being more destructive than softer photons. If the initial PAH abundance does not depend on metallicity, while their destruction rate depends on metallicity, we would expect the 12~+~log(O/H)--$F_8/F_{24}$ correlation to be weaker (or absent) in younger complexes and to become stronger in older complexes. Also, we would expect the $F_8/F_{24}$ ratio to become smaller with time if PAHs are destroyed more effectively than larger grains \citep{allain}. Our complexes have sizes of the order of a few hundred pc, which is much bigger than the size of an individual H~{\sc ii} region. However, this does not mean that processes we discuss operate on this spatial scale. What we observe is actually a merged result of dust evolution in many separate regions, which are, probably, observed in our Galaxy as infrared bubbles \citep{bubbles}.

A correlation between the $F_{8}/F_{24}$ ratio and metallicity for our entire sample is presented in Fig.~\ref{ohf824}a. Metallicity error bars do not include the intrinsic uncertainty of the ONS method which is of the order of 0.08 \citep{ONS}. For extra clarity points are marked with different colours from blue for low metallicity to red for high metallicity. Having the `destructive' scenario in mind, we may also want to check whether $F_{8}/F_{24}$ correlates with the [O~{\sc iii}]$\lambda 5007/\mathrm{H\beta}$ ratio, which is the only (albeit poor) UV field hardness parameter that we have at hand. The correlation does exist as Fig.~\ref{ohf824}b demonstrates but circle colours, showing metallicity, indicate that it may be a mere reflection of the correlation between [O~{\sc iii}]$\lambda 5007/\mathrm{H\beta}$ and 12~+~log(O/H).

\begin{figure*}
\includegraphics[width=0.9\hdsize]{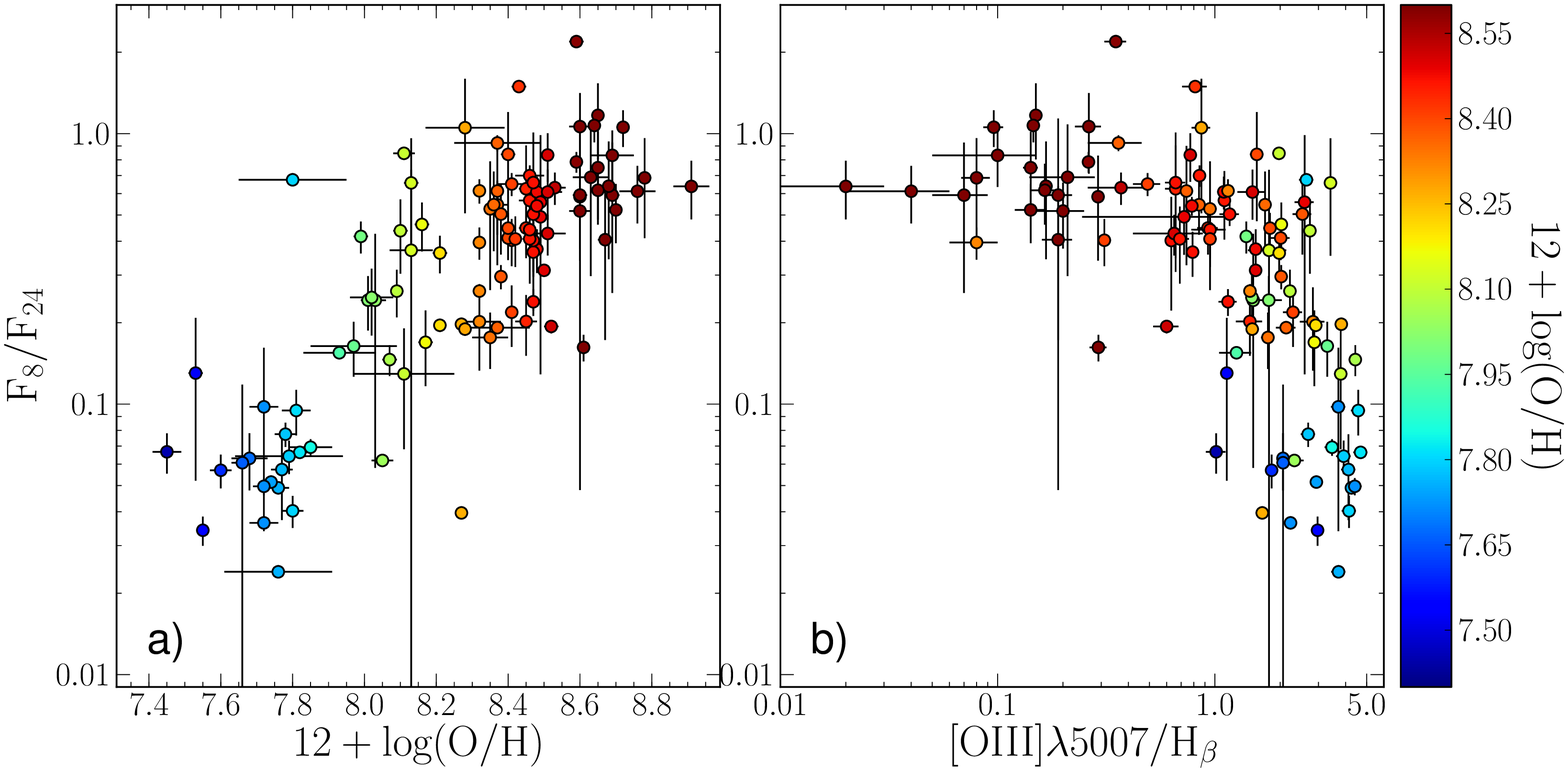}
\caption{a) The $F_{8}/F_{24}$ ratio as a function of metallicity in the studied galaxies. Metallicity values are also indicated by colours from blue for low values to violet for high values. b) Relation between the [O~{\sc iii}]$\lambda 5007/\mathrm{H\beta}$ ratio which is often used as an indicator of the radiation field hardness and the infrared flux ratio.}
\label{ohf824}
\end{figure*}

Apparently, the correlation between metallicity and the PAH content exists when we consider complexes of all ages. Further, we divided our complexes into five age bins, so that each bin contains about 20 complexes. In Fig.~\ref{difage} we show correlations between the infrared flux ratio and metallicity for each age bin separately. The corresponding Spearman rank correlation coefficients $r_{\rm S}$ are indicated on each panel together with the number $N$ of standard deviations by which the sum squared difference of ranks deviates from its value for the null hypothesis \citep{numrec}. The correlation is strong, with $r_{\rm S}\sim0.7-0.8$, at all ages. So, contrary to our na\"ive expectations, the correlation between metallicity and the PAH content (or, at least, the $F_8/F_{24}$ ratio) does not get any stronger with time.

\begin{figure*}
\includegraphics[width=0.6\hdsize]{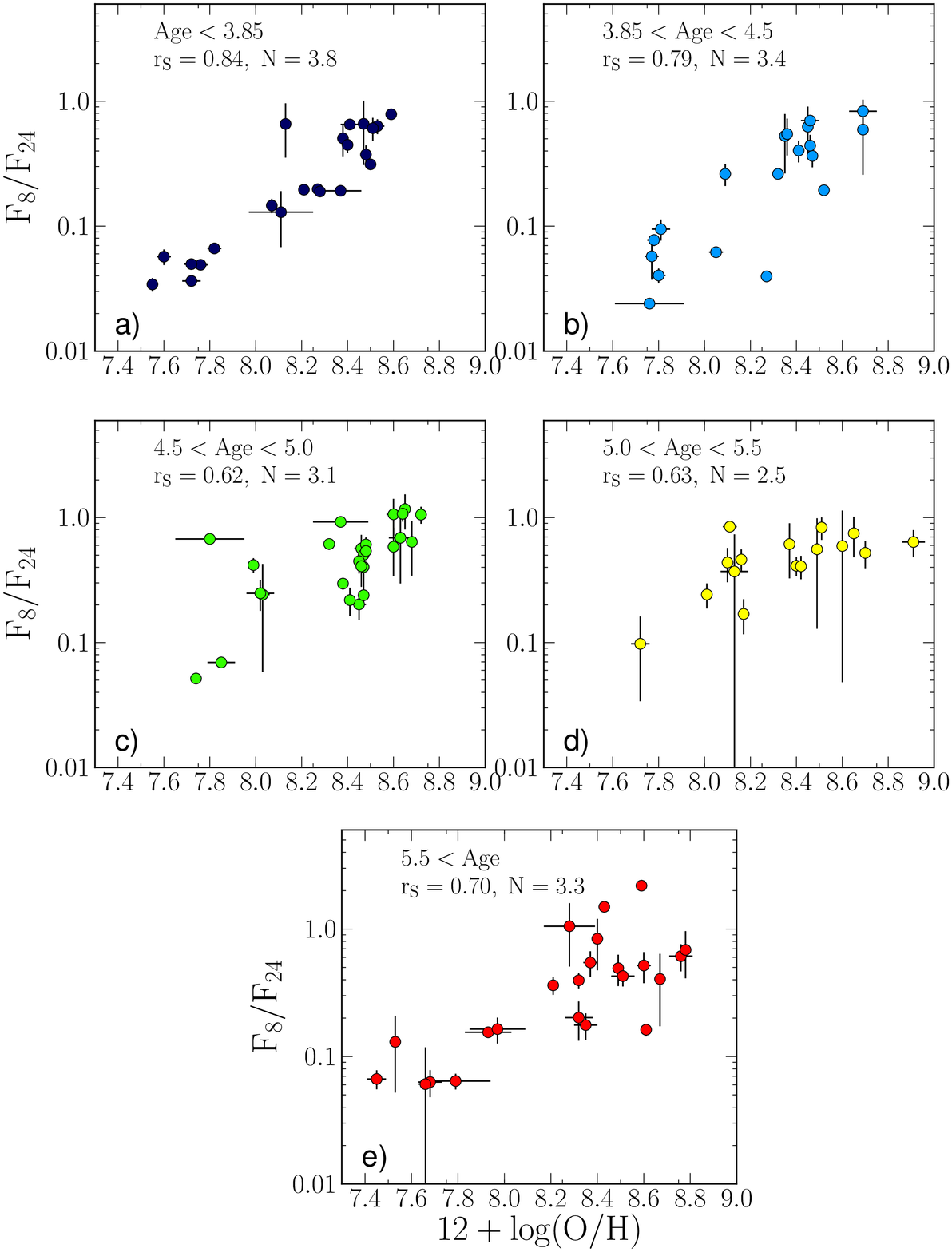}
\caption{Relation between the infrared flux ratio and the metallicity of the complex for five age bins. Limits for bins are indicated on each panel along with the Spearman rank correlation coefficients $r_{\rm S}$ and the number $N$ of standard deviations by which the sum squared difference of ranks deviates from its value for the null hypothesis. Various colours are used to denote the H~{\sc ii} complex metallicity as indicated by the scale in Fig.~\ref{ohf824}.}
\label{difage}
\end{figure*}

Next we present a correlation between the $F_{8}/F_{24}$ ratio and age, and here results are even more unexpected. We have divided all our H~{\sc ii} complexes into five metallicity groups, again with each group containing about 20 complexes. In Fig.~\ref{difz} we plot $F_{8}/F_{24}$ vs age for all these groups, indicating their metallicity limits, $r_{\rm S}$ and $N$. All the complexes are considered, with the one exception. 

\begin{figure*}
\includegraphics[width=0.6\hdsize]{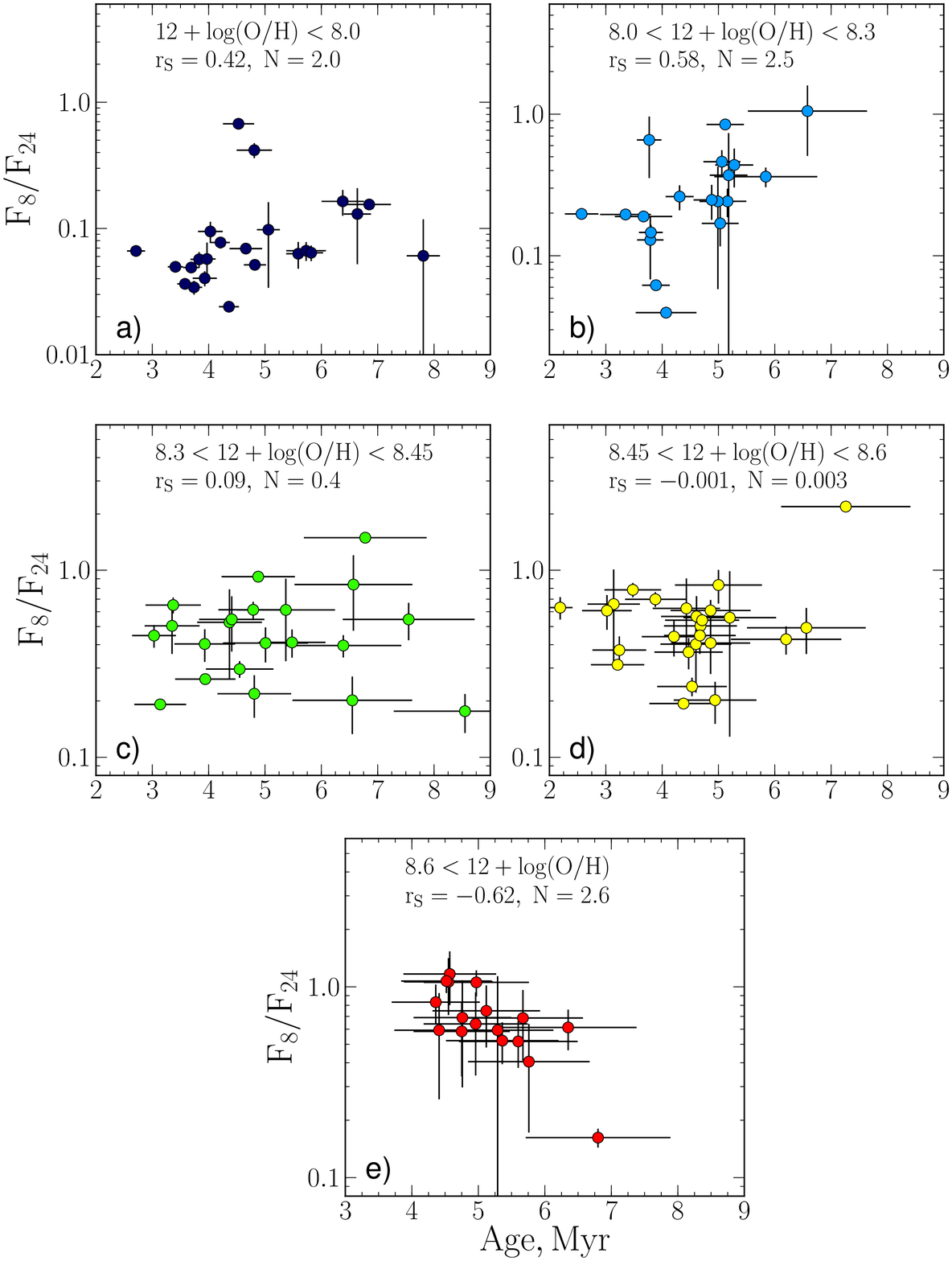}
\caption{Relation between the infrared flux ratio and the age of the complex for five metallicity bins. Limits for bins are indicated on each panel along with the Spearman rank correlation coefficients. Various colours are used to denote the H~{\sc ii} complex metallicity as indicated by the scale in Fig.~\ref{ohf824}.}
\label{difz}
\end{figure*}

The correlation for the lowest metallicity bin is not very impressive, with $r_{\rm S}=0.42$. However, it is somewhat spoiled by two outliers. One of them is HSK~61 region in Holmberg~II (the rightmost point in top left panel of Fig.~\ref{difz}) for which significantly smaller age has been inferred by \cite{Stewart_ho} (see Fig.~\ref{comparing}). Another outlier is the IC~2574-11 region that has an order of magnitude higher $F_{8}/F_{24}$ ratio than all the other complexes in this galaxy and is visible as an outlier in Fig.~\ref{ohf824} at metallicity of 7.8. This region can be associated with the supernova remnant (\cite{walter98}, O.V. Egorov et al., 2014). Without these two regions $r_{\rm S}$ would be 0.51 for this metallicity range.

For a higher metallicity, between 8.0 and 8.3, the correlation coefficient is somewhat higher and it is again positive, corresponding to the $F_{8}/F_{24}$ ratio growing with age. At even higher metallicity, from 8.3 to 8.45, the correlation becomes almost non-existent. There is again an outlier, NGC~7741-9, the oldest region in our sample. Without it $r_{\rm S}$ would be somewhat higher, 0.26 instead of 0.09. For 12~+~log(O/H) values between 8.45 and 8.6 the correlation disappears completely.

Finally, for metallicities above 8.6 the correlation arises again, but with the opposite sign. With $r_{\rm S}=-0.62$, it indicates that the $F_{8}/F_{24}$ ratio decreases with age. The oldest region in this subsample, +074+064 in NGC~3184, helps to stress the correlation visually, but does not play a defining role. Without this region $r_{\rm S}$ would be --0.55 in this metallicity range.

Again, contrary to our expectation, the $F_{8}/F_{24}$ ratio becomes smaller with time only in complexes with high metallicities. At lower metallicities it actually grows. It does not imply, of course, that mass of PAHs in the complex grows, but indicates that particles emitting at 8\,$\mu$m become {\em relatively\/} more abundant with time than particles emitting at 24\,$\mu$m.

Our ability to test the connection between the $F_{8}/F_{24}$ ratio and the hard-to-soft flux ratio is limited as [O~{\sc iii}]$\lambda 5007/\mathrm{H\beta}$ is actually not a good tracer of the radiation field hardness. We have checked whether the $F_{8}/F_{24}$ ratio correlates with the [O~{\sc iii}]$\lambda 5007/\mathrm{H\beta}$ ratio in separate metallicity bins and found a weak negative correlation in all bins, with $r_{\rm S}$ of the order of --0.3. The significance of this correlation is hard to ascertain. Even if [O~{\sc iii}]$\lambda 5007/\mathrm{H\beta}$ were a reliable tracer of the radiation field hardness, it still depends both on age and metallicity. To estimate the role of the radiation hardness directly, we would need to consider groups of H~{\sc ii} complexes having similar (preferably low) metallicities {\em and\/} ages, provided the radiation hardness within the group varies for some reason, different from metallicity and age. However, our sample is definitely too small for this task.

The summary of our findings is as follows. The mid-infrared flux ratio grows with metallicity at all ages. The negative correlation of the ratio with the [O~{\sc iii}]$\lambda 5007/\mathrm{H\beta}$ mostly reflects the correlation between this ratio and metallicity. When we consider separate metallicity bins, only weak remaining correlation between $F_{8}/F_{24}$ and [O~{\sc iii}]$\lambda 5007/\mathrm{H\beta}$ is observed. The most interesting result is that the $F_{8}/F_{24}$ ratio grows with age at low metallicities and decreases with age at high metallicities.

\section{Discussion}

The main problem we tried to address in this study is the still unexplained origin of the correlation between metallicity and PAH emission (or abundance) shown in Fig.~\ref{ohf824} for our sample. This problem was discussed by many authors, and a number of possible explanations were put forward. However, none of these explanations is widely accepted.

One of the hypotheses states that PAHs are more efficiently destroyed in low metallicity environments due to factors intrinsic to such environments. For example, supernova shocks have been invoked as a factor that can rapidly dissociate small particles at low metallicity \citep{ohalloran}. However, \citet{Sandstrom} did not find any correlation between PAHs and supernova shocks in the SMC.

Another plausible destruction factor is the UV field. Using the [Ne~{\sc iii}]/[Ne~{\sc ii}] line ratio as a tracer of the radiation field hardness, \citet{Madden2000} showed that starlight and nebular spectrum is harder in low metallicity medium than in high metallicity medium. In principle, it can be expected that harder UV field, rich in more energetic photons, dissociates PAH particles faster. \citet{Khramtsova} presented arguments in favour of the `destructive' scenario of PAH evolution, relating the lower PAH mass fraction in low-metallicity environments to a stronger and/or harder UV field. In this study we approach this problem using an H~{\sc ii} complex sample that encompasses a wide metallicity range. Having estimated ages of the H~{\sc ii} complexes, we may try to reveal some evolutionary effects in the PAH content of H~{\sc ii} complexes.

Let us first suppose that the only process that regulates the relative PAHs content in H~{\sc ii} complexes is their destruction by UV photons. In this case we would expect to see $q_{\rm PAH}$ and the $F_{8}/F_{24}$ ratio decreasing with age at all metallicities. Also, if the photodestruction efficiency decreases with metallicity, a correlation of the PAH tracers with 12~+~log(O/H) should become stronger with age. None of these predictions is confirmed by our data. The correlation between the $F_{8}/F_{24}$ ratio and metallicity is already strong in the youngest of our complexes and does not change much with time. The negative correlation of the $F_{8}/F_{24}$ ratio with age is only seen in the highest metallicity complexes, while in the complexes with low metal content the $F_{8}/F_{24}$ ratio actually increases with time.

Apparently, the real picture is not limited to just a PAH photodestruction in H~{\sc ii} complexes. In particular, the growing $F_{8}/F_{24}$ ratio does not necessarily mean that $q_{\rm PAH}$ increases. The $F_{8}/F_{24}$ ratio could also grow in these complexes as a result of more efficient destruction of VSGs \citep{Pilleri}. The destruction of larger grains means that the grain size distribution evolves and, strictly speaking, the \cite{DL07} model with the size distribution taken from \cite{weingartner}, which we invoked to justify usage of the $F_{8}/F_{24}$ ratio as a measure of the relative PAH content, may no longer be applicable, and different size distributions should be used. That is, in principle, we can imagine the situation when $F_{8}/F_{24}$ is large and $q_{\rm PAH}$ is small.

Then, VSGs may not be simply eliminated. Their destruction can be a gradual process, producing PAHs in its course. If VSGs are indeed precursors of PAHs, the PAH relative abundance and the $F_{8}/F_{24}$ ratio are determined by a more delicate balance between VSG and PAH photodestruction timescales. For example, if PAHs are more stable to photodestruction than VSGs \citep{Pilleri}, their respective abundances would both decrease with time, but abundances of VSGs would drop faster. Also, PAHs formed as a result of VSGs destruction would partially compensate for the destruction of original PAHs.

To explain the observed correlation between the metallicity and the PAH tracers, we might have assumed, for example, that the efficiency of VSG destruction is more sensitive to the UV field hardness than the efficiency of PAH destruction. But this assumption has to be reconciled with the lack of correlation between metallicity and the ratio of the 24\,$\mu$m flux and the flux at longer wavelengths (the $P_{24}$ index introduced in the work of \cite{DL07} that was shown in the work of \cite{Khramtsova}. It implies that VSG destruction occurs irrespective to metallicity and the observed change in the evolution of the $F_{8}/F_{24}$ ratio from low to high metallicity is related solely to the PAH evolution.

It must be admitted that the original question on the nature of the correlation between the metallicity and the $F_{8}/F_{24}$ ratio (or $q_{\rm PAH}$) still remains unanswered. Our data show that the correlation already exists in complexes having ages less than 3~Myr (Fig.~\ref{difage}). So, whatever mechanisms are responsible for this correlation, they are already operative at the early stages of H~{\sc ii} complexes evolution or prior to these stages. These mechanisms can be related to the specific details of dust destruction in the youngest H~{\sc ii} complexes, or to the details of dust evolution in molecular clouds.

The idea that PAHs and grains of other types can form in molecular clouds is by no means new. Some mechanisms of PAH growth in the dense ISM were suggested as an addition to stellar sources \citep{Greenberg,Parker11}. There are some observational indications that the PAH abundance increases in dense molecular clouds. \citet{Sandstrom} found the correlation between CO and the PAH dust mass fraction in the Small Magellanic Cloud while the correlation of $q_{\rm PAH}$ with the distribution of AGB stars is not observed. They concluded that perhaps those PAHs that we observe in bright H~{\sc ii} complexes were formed in cold molecular clouds directly prior to the onset of the star formation, while PAHs ejected by stars were destroyed in the diffuse ISM. If efficiency of PAH formation in molecular clouds depend on metallicity, e.g., through its relation to surface chemistry, we may expect lower PAH content in lower metallicity environment.

To draw more specific conclusions, one would need to consider H~{\sc ii} complexes with both optical spectroscopy and far-infrared photometry available. In this study we only use {\em Spitzer\/} photometry as it allows considering smaller complexes and complexes in crowded environments. Most of the galaxies from our sample have also been observed with {\em Herschel\/} but these data would not help to clarify the situation as they allow studying only large and/or isolated regions. In our sample only H~{\sc ii} complexes in Holmberg II correspond to these criteria. Their study is published separately \citep{wiebe14}. For other galaxies far infrared and submillimetre data of higher angular resolution are needed, possibly, accessible with the Atacama Large Millimeter Array (ALMA).

\section{Summary}

Analysis of optical spectroscopy and infrared photometry of H~{\sc ii} complexes in nine galaxies is presented. Observational data are used to infer metallicities, ages, and the radiation field hardness in the studied complexes and to estimate the ratio of fluxes at 8\,$\mu$m and 24\,$\mu$m considered as an indicator of PAH abundance.

We use the Starburst99 code to compute theoretical spectra for our regions and the grid of theoretical line ratios provided by \cite{2010AJ....139..712L} to assess the reliability of [S~{\sc ii}]/H$\alpha$, [O~{\sc iii}]$\lambda 5007/\mathrm{H\beta}$, and [Ne~{\sc iii}]15\,$\mu$m/[Ne~{\sc ii}]12\,$\mu$m line ratios as tracers of the radiation field hardness. This quantity is evaluated as a ratio of integrated fluxes in the ranges from 90\AA\ to 912\AA\ and from 912\AA\ to 2000\AA\ (`hard-to-soft flux ratio'). The [S~{\sc ii}]/H$\alpha$ ratio only depends on the hard-to-soft flux ratio if the ionization parameter, $q$, is relatively large, $\ga10^8$~cm s$^{-1}$. At smaller values of $q$ the [S~{\sc ii}]/H$\alpha$ ratio does not depend on the hardness of radiation field. Oxygen and neon line ratios are more sensitive to the hard-to-soft flux ratio and can be used as tracers of radiation field hardness, albeit with uncertainty if the ionization parameter is unknown.

Our main conclusions are the following:
\begin{itemize}
\item{As found in previous studies, the $F_{8}/F_{24}$ ratio is an increasing function of metallicity in the studied complexes. The [O~{\sc iii}]$\lambda 5007/\mathrm{H\beta}$ line ratio decreases with metallicity, in quantitative agreement with the numerical models. The infrared flux ratio is also correlated with the [O~{\sc iii}]$\lambda 5007/\mathrm{H\beta}$ but the available data are not sufficient to make definitive conclusions about radiation hardness being the driver of the correlation between metallicity and the PAH tracers. In separate metallicity bins the correlation between the $F_{8}/F_{24}$ ratio and the radiation hardness is much weaker.}
\item{The relative abundance of PAHs (measured with the $F_{8}/F_{24}$ ratio) grows with age of an H~{\sc ii} complex at low metallicities (less than $\sim8.3$) and decreases with age in H~{\sc ii} complexes with metallicities higher than about 8.6. The correlation between the $F_{8}/F_{24}$ ratio and metallicity is already present in the youngest complexes from our sample (with ages less than 3~Myr) and does not change appreciably over the following evolution.}\end{itemize}

The abundance of PAHs in different physical environments is defined by a number of competing processes that can be metallicity-dependent, age-dependent, etc. In order to understand the connection between the PAH abundance and the star formation rate as well as other parameters of star-forming regions, all the processes need to be considered that influence the abundance of PAHs including formation of PAHs in stars, molecular clouds, and H~{\sc ii} complexes as well as their destruction by UV photons, shock waves, cosmic rays, shattering etc. All these processes can be equally important in the PAH lifecycle and we still need to find out which of them play the important role and which can be neglected. This study represents yet another step in solving the problem of the PAH evolution and put some additional constraints on various factors that affect this evolution.

\section*{Acknowledgments}
We thank the anonymous referee for criticism and suggestions that led to significant improvements of the paper, Kevin Croxall for spectroscopic data on H~{\sc ii} complexes in IC~2574 and Emily Levesque for valuable advices. This work was supported by the RFBR grants 12-02-31356, 14-02-31456, 14-02-00604 and also the OFN-17 Research Program of the Division of Physics, Russian Academy of Sciences. This work is based in part on observations obtained with 6-m telescope of the Special Astrophysical Observatory of Russian Academy of Sciences, and in part on observations made with the Spitzer Space Telescope, which is operated by the Jet Propulsion Laboratory, California Institute of Technology under a contract with NASA. We used the HyperLeda database (http://leda.univ-lyon1.fr).

\bibliographystyle{mn2e} 
\bibliography{refs}

\onecolumn \tiny
\begin{landscape}
\begin{longtable}{lllllllllllll}
\caption{List of studied H~{\sc ii} complexes in galaxies IC~1727 and NGC~7741 and results of spectroscopy of H~{\sc ii} complexes. The table contains designations of complexes, equatorial coordinates, bright line intensities in studied H~{\sc ii} complexes normalised to H$\beta$, equivalent width of the H$\beta$ line, metallicity estimated by the `ONS' method and radial distance from the centre of a galaxy normalised to a R$_{25}$.}\\ \hline
\label{table:lines}
Complex   &  $\alpha$ offset &  $\delta$ offset & I([O~{\sc ii}])& I([O~{\sc iii}]) & I([O~{\sc iii}]) & I(H$\alpha$) & 
I([N~{\sc ii}]) & I([S~{\sc ii}]) &EW(H$\beta$) &E(B-V)&$\mathrm{12+log(O/H)}$ &$r/R_{25}$\\ 
& $^{\arcsec}$& $^{\arcsec}$ & $\lambda3727+\lambda3729$& $\lambda4959$ &
$\lambda5007$ & & $\lambda6548+\lambda6584$ &$\lambda6717+\lambda6731$ & \AA & &ONS&\\
\hline
\multicolumn {13}{l}{IC~1727}   \\ \hline
IC~1727-1&$+1.68$&$-29.5$    & 2.57 $\pm$  0.67 &	 0.63 $\pm$  0.08 &  1.88 $\pm$  0.06 &	 3.07 $\pm$  0.15 &  0.41 $\pm$  0.17 &	 1.14 $\pm$  0.18& 103.9	$\pm$ 0.6 &
0.33$\pm$0.03&8.07 $\pm$ 0.03 &0.19\\
IC~1727-2&$+1.61$&$-25.0$    & 2.85 $\pm$  0.23 &	 1.02 $\pm$  0.03 &  3.13 $\pm$  0.03 &	 2.85 $\pm$  0.05 &  0.18 $\pm$  0.07 &	 0.47 $\pm$  0.09& 91.8 $\pm$ 0.4  &
0.29$\pm$0.03&8.02 $\pm$ 0.04&0.18\\
IC~1727-3&$+1.23$&$-18.9$    & 2.85 $\pm$  0.23 &	 1.02 $\pm$  0.03 &  3.13 $\pm$  0.03 &	 2.85 $\pm$  0.05 &  0.18 $\pm$  0.07 &	 0.47 $\pm$  0.09& 44.9 $\pm$ 0.1 &
0.33$\pm$0.02&8.04 $\pm$ 0.06&0.14\\
IC~1727-4&$+1.10$&$-16.5$    & 3.87 $\pm$  0.31 &	 0.50 $\pm$  0.03 &  1.51 $\pm$  0.03 &	 2.85 $\pm$  0.07 &  0.32 $\pm$  0.10 &	 0.69 $\pm$  0.09& 51.1 $\pm$ 0.3&
0.22$\pm$0.02&8.03 $\pm $ 0.04 &0.12\\
IC~1727-5&$+0.59$&$-5.0$	& 3.86 $\pm$  0.18 &	 0.51 $\pm$  0.02 &  1.53 $\pm$  0.03 &	 2.87 $\pm$  0.05 &  0.33 $\pm$  0.07 &	 0.71 $\pm$  0.05& 52.7 $\pm$ 0.3&
0.24$\pm$0.02&8.04 $\pm$ 0.03 &0.41\\
IC~1727-6&$+2.17$&$-29.3$	& 3.59 $\pm$  0.20 &	 0.52 $\pm$  0.02 &  1.47 $\pm$  0.03 &	 2.92 $\pm$  0.06 &  0.26 $\pm$  0.11 &	 0.90 $\pm$  0.07& 60.2 $\pm$ 0.5&
0.33$\pm$0.04&8.05 $\pm$ 0.05&0.26\\
IC~1727-7&$+1.95$&$-26.2$	& 4.22 $\pm$  0.25 &	 0.45 $\pm$  0.02 &  1.38 $\pm$  0.03 &	 2.72 $\pm$  0.05 &  0.26 $\pm$  0.06 &	 0.78 $\pm$  0.06& 25.1	$\pm$ 0.1&
0.36$\pm$0.04& 8.04 $\pm$ 0.04&0.24\\
IC~1727-8&$+1.77$&$-17.5$	& 2.34 $\pm$  0.11 &	 1.43 $\pm$  0.01 &  4.20 $\pm$  0.01 &	 2.88 $\pm$  0.02 &  0.17 $\pm$  0.02 &	 0.52 $\pm$  0.03& 55.4	$\pm$ 0.1&
0.34$\pm$0.15& 8.07 $\pm$ 0.03&0.22\\
IC~1727-9&$+0.56$&$+13.6$5	& 4.44 $\pm$  0.66 &	 0.41 $\pm$  0.06 &  1.09 $\pm$  0.07 &	 3.02 $\pm$  0.10 &  0.24 $\pm$  0.14 &	 0.82 $\pm$  0.37& 22.1 $\pm$ 1.1&
0.32$\pm$0.05& 7.94 $\pm$ 0.09 &0.13\\
IC~1727-10&$-5.92$&$+81.4$	& 5.18 $\pm$  0.39 &	 0.72 $\pm$  0.03 &  2.12 $\pm$  0.04 &	 2.88 $\pm$  0.03 &  0.23 $\pm$  0.04 &	 0.46 $\pm$  0.07& 120.0 $\pm$ 1.1&
0.06$\pm$0.03& 8.05 $\pm$ 0.04&0.69\\
IC~1727-11&$-6.49$&$+110.0$	& 2.57 $\pm$  1.12 &	 1.24 $\pm$  0.11 &  3.86 $\pm$  0.11 &	 2.86 $\pm$  0.13 &  0.28 $\pm$  0.22 &	 0.56 $\pm$  0.30& 129.3 $\pm$ 2.5&
0.06$\pm$0.03& 8.00 $\pm$ 0.03&0.79\\
\hline
\multicolumn {13}{c}{NGC~7741}   \\ \hline
NGC~7741-1&$+3.45$&$+1.8$	& 2.44 $\pm$  0.72 &	 0.59 $\pm$  0.07 &  1.79 $\pm$  0.06 &	 2.93 $\pm$  0.07 &  0.61 $\pm$  0.09 &	 1.12 $\pm$  0.11& 44.8	$\pm$ 0.8&0.23$\pm$0.05
&8.13 $\pm$ 0.13 &0.62\\
NGC~7741-2&$+0.68$&$+1.6$	& 3.69 $\pm$  0.82 &	 0.39 $\pm$  0.08 &  1.12 $\pm$  0.09 &	 2.78 $\pm$  0.12 &  0.77 $\pm$  0.14 &	 1.12 $\pm$  0.24& 4.8 $\pm$	0.2&0.34$\pm$0.19
&8.47 $\pm$ 0.04 &0.12\\
NGC~7741-3&$-0.72$&$+5.4$	& 2.72 $\pm$  0.14 &	 0.19 $\pm$  0.03 &  0.62 $\pm$  0.02 &	 2.79 $\pm$  0.03 &  0.82 $\pm$  0.05 &	 0.81 $\pm$  0.05& 13.7	$\pm$ 0.2&0.34$\pm$0.03
&8.47 $\pm$ 0.11 &0.13\\
NGC~7741-4&$-1.29$&$+5.4$	& 3.48 $\pm$  0.26 &	 0.20 $\pm$  0.03 &  0.71 $\pm$  0.03 &	 2.84 $\pm$  0.03 &  0.67 $\pm$  0.05 &	 0.71 $\pm$  0.07& 8.6	$\pm$ 0.1&0.39$\pm$0.06
&8.39 $\pm$ 0.06 &0.23\\
NGC~7741-5&$-1.61$&$+5.4$	& 2.35 $\pm$  0.25 &	 0.28 $\pm$  0.04 &  0.88 $\pm$  0.04 &	 2.87 $\pm$  0.05 &  0.71 $\pm$  0.08 &	 0.66 $\pm$  0.13& 18.7	$\pm$ 0.2&0.18$\pm$0.03
&8.44 $\pm$ 0.05 &0.28\\
NGC~7741-6&$-1.90$&$+5.9$	& 4.43 $\pm$  0.88 &	 0.36 $\pm$  0.08 &  1.04 $\pm$  0.10 &	 2.82 $\pm$  0.14 &  0.58 $\pm$  0.19 &	 0.94 $\pm$  0.24& 12.1	$\pm$ 0.4&0.20$\pm$0.07
&8.30 $\pm$ 0.05&0.34\\
NGC~7741-7&$-3.76$&$+10.7$	& 3.21 $\pm$  1.14 &	 0.25 $\pm$  0.09 &  0.80 $\pm$  0.09 &	 2.84 $\pm$  0.10 &  0.50 $\pm$  0.13 &	 0.73 $\pm$  0.37& 13.5	$\pm$ 0.4&0.26$\pm$0.09
&8.29 $\pm$ 0.06&0.66\\
NGC~7741-8&$+2.18$&$-0.8$	& 2.69 $\pm$  0.60 &	 0.56 $\pm$  0.06 &  1.60 $\pm$  0.06 &	 2.87 $\pm$  0.06 &  0.48 $\pm$  0.09 &	 0.68 $\pm$  0.23& 47.0	$\pm$ 1.0&0.04$\pm$0.03
&8.33 $\pm$ 0.04&0.40\\
NGC~7741-9&$+1.93$&$+11.7$	& 2.24 $\pm$  0.26 &	 0.64 $\pm$  0.02 &  1.77 $\pm$  0.02 &	 2.85 $\pm$  0.03 &  0.47 $\pm$  0.06 &	 0.50 $\pm$  0.10& 5.5	$\pm$ 0.1&0.13$\pm$0.03
&8.35 $\pm$ 0.03&0.39\\
NGC~7741-10&$+1.78$&$+19.9$	& 2.27 $\pm$  0.32 &	 0.94 $\pm$  0.01 &  2.81 $\pm$  0.01 &	 2.84 $\pm$  0.04 &  0.38 $\pm$  0.05 &	 0.42 $\pm$  0.12& 13.3	$\pm$ 0.1&0.07$\pm$0.01
&8.30 $\pm$ 0.03&0.42\\
\hline
\end{longtable}
\end{landscape}

\onecolumn
\begin{landscape}
    \begin{longtable}{lllllllllll}
\caption{Results of aperture infrared photometry of H~{\sc ii} complexes including designation of a complex, equatorial coordinates, aperture size, and fluxes at 3.6, 4.5, 5.8, 8.0 and 24~$\mu$m.}\\

\hline
Complex &  $\alpha$(J2000)  &  $\delta$(J2000)  &  Aperture  &  $F_{3.6}$, mJy &  $F_{4.5}$, 
mJy & $F_{5.8}$, mJy & $F_{8.0}$, mJy &$F_{8.0}^{\rm afe}$, mJy & $F_{24}$, mJy & $F_{24}^{\rm ns}$, mJy \\ \hline
& $^{h}\quad ^{m}\quad ^{s}$ &  $^\circ \quad \arcmin \quad \arcsec$         & radius, $\arcsec$&           &          &          &       &      \\ 
\hline \endhead

\multicolumn {10}{c}{IC~1727}   \\ \hline
IC~1727-3 & 01 47 31.14 & +27 19 38.39 & 8.0 & 0.237 $\pm$ 0.032 & 0.241
$\pm$ 0.032 & 0.516 $\pm$ 0.051 & 0.782 $\pm$ 0.097 & 0.488 $\pm$ 0.094 & 2.100 $\pm$ 0.242 & 2.013 $\pm$ 0.242\\
IC~1727-5 & 01 47 30.44 & +27 19 52.51 & 8.0 & 0.152 $\pm$ 0.026 & 0.154
$\pm$ 0.027 & 0.155 $\pm$ 0.041 & 0.216 $\pm$ 0.061 & 0.095 $\pm$ 0.060 & 0.448 $\pm$ 0.163 & 0.392 $\pm$ 0.163\\
IC~1727-6 & 01 47 32.09 & +27 19 28.72 & 8.0 & 0.141 $\pm$ 0.033 & 0.144
$\pm$ 0.033 & 0.346 $\pm$ 0.060 & 0.627 $\pm$ 0.102 & 0.426 $\pm$ 0.098 & 1.770 $\pm$ 0.263 & 1.718 $\pm$ 0.263\\
IC~1727-9 & 01 47 30.46 & +27 19 11.38 & 8.0 & 0.084 $\pm$ 0.035 & 0.086
$\pm$ 0.035 & 0.199 $\pm$ 0.046 & 0.131 $\pm$ 0.054 & 0.042 $\pm$ 0.055 & 0.303 $\pm$ 0.095 & 0.272 $\pm$ 0.096\\
IC~1727-10 & 01 47 24.00 & +27 19 19.00 & 8.0 & 0.011 $\pm$ 0.011 & 0.011
$\pm$ 0.011 & 0.002 $\pm$ 0.018 & 0.024 $\pm$ 0.023 & 0.017 $\pm$ 0.024 & 0.279 $\pm$ 0.056 & 0.0275 $\pm$ 0.056\\
IC~1727-11 & 01 47 23.41 & +27 19 49.98 & 8.0 & 0.016 $\pm$ 0.008 & 0.016
$\pm$ 0.008 & 0.040 $\pm$ 0.013 & 0.069 $\pm$ 0.022 & 0.048 $\pm$ 0.022 & 0.379 $\pm$ 0.059 & 0.373 $\pm$ 0.059\\\hline
\multicolumn {10}{c}{NGC~7741}   \\ \hline
NGC~7741-1 & 23 43 58.462 & +26 04 21.95 & 6.0 & 0.0134 $\pm$ 0.027 & 0.007 $\pm$ 0.017 & 0.043 $\pm$ 0.046 & 0.133 $\pm$ 0.103 & 0.107 $\pm$ 0.097 & 0.290 $\pm$ 0.108 & 0.290 $\pm$ 0.108   \\ 
NGC~7741-3 & 23 43 54.343 & +26 04 34.94 & 6.0 & 0.958 $\pm$ 0.204 & 0.664 $\pm$ 0.138 & 2.240 $\pm$ 0.403 & 5.070 $\pm$ 0.878 & 3.754 $\pm$ 0.843 & 7.640 $\pm$ 1.220 & 7.623 $\pm$ 1.220\\
NGC~7741-4 & 23 43 53.695 & +26 04 33.21 & 6.0 & 1.530 $\pm$ 0.202 & 1.050 $\pm$ 0.140 & 3.560 $\pm$ 0.442 & 7.980 $\pm$ 1.050 & 5.826 $\pm$ 1.015 & 10.700 $\pm$ 1.540 & 10.670 $\pm$ 1.540\\
NGC~7741-7 & 23 43 52.301 & +26 04 29.74 & 6.0 & 0.235 $\pm$ 0.083 & 0.154 $\pm$ 0.055 & 0.739 $\pm$ 0.186 & 1.850 $\pm$ 0.440 & 1.247 $\pm$ 0.456 & 1.190 $\pm$ 0.434 & 1.186 $\pm$ 0.434\\
NGC~7741-9 & 23 43 56.877 & +26 04 39.74 & 6.0 & 0.222 $\pm$ 0.058 & 0.227 $\pm$ 0.043 & 0.791 $\pm$ 0.144 & 1.810 $\pm$ 0.312 & 1.426 $\pm$ 0.294 & 8.090 $\pm$ 0.919 & 8.086 $\pm$ 0.919\\
NGC~7741-10 & 23 43 56.719 & +26 04 47.64 & 6.0 & 0.111 $\pm$ 0.052 & 0.103 $\pm$ 0.043 & 0.592 $\pm$ 0.132& 1.330 $\pm$ 0.283 & 1.052 $\pm$ 0.267 & 5.220 $\pm$ 1.180 & 5.218 $\pm$ 1.180\\\hline
\multicolumn {10}{c}{Holmberg II}   \\ \hline
HSK~7 & 08 18 49.80 & +70 44 48.92 & 10.0 & 0.225 $\pm$  0.008 &  0.277 $\pm$  0.007 &  0.472 $\pm$  0.008 &  0.773 $\pm$  0.014 &  0.503 $\pm$  0.014 & 10.200 $\pm$  0.118 & 10.118 $\pm$  0.118  \\
HSK~15 & 08 18 54.92 & +70 43 11.34 & 8.0 & 0.142 $\pm$  0.019 &  0.097 $\pm$  0.014 &  0.097 $\pm$  0.017 &  0.075 $\pm$  0.016 &  0.001 $\pm$  0.017 &  0.781 $\pm$  0.077 &  0.730 $\pm$  0.078  \\
HSK~16-17 & 08 18 54.92 & +70 42 57.95 & 8.0 & 0.083 $\pm$  0.026 &  0.079 $\pm$  0.019 &  0.118 $\pm$  0.012 &  0.156 $\pm$  0.015 &  0.086 $\pm$  0.019 &  1.350 $\pm$  0.052 &  1.320 $\pm$  0.053  \\
HSK~20 & 08 18 56.27 & +70 42 38.27 & 10.0 & 0.061 $\pm$  0.010 &  0.036 $\pm$  0.008 &  0.107 $\pm$  0.009 &  0.118 $\pm$  0.009 &  0.066 $\pm$  0.010 &  0.989 $\pm$  0.070 &  0.967 $\pm$  0.070  \\
HSK~25 & 08 18 59.90 & +70 43 01.02 & 6.0 & 0.029 $\pm$  0.019 &  0.018 $\pm$  0.015 &  0.056 $\pm$  0.013 &  0.073 $\pm$  0.024 &  0.045 $\pm$  0.025 &  0.353 $\pm$  0.074 &  0.343 $\pm$  0.075  \\
HSK~31 & 08 19 02.66 & +70 42 59.41 & 10.0 & 0.007 $\pm$  0.014 &  0.004 $\pm$  0.010 &  0.059 $\pm$  0.021 &  0.066 $\pm$  0.021 &  0.046 $\pm$  0.021 &  0.474 $\pm$  0.069 &  0.471 $\pm$  0.070  \\
HSK~59 & 08 19 23.66 & +70 42 58.95 & 8.0 & 0.001 $\pm$  0.010 &  0.005 $\pm$  0.007 &  0.023 $\pm$  0.005 &  0.107 $\pm$  0.011 &  0.091 $\pm$  0.011 &  1.590 $\pm$  0.096 &  1.590 $\pm$  0.096  \\
HSK~61 & 08 19 23.13 & +70 41 54.07 & 10.0 & 0.053 $\pm$  0.010 &  0.051 $\pm$  0.009 &  0.053 $\pm$  0.016 &  0.051 $\pm$  0.010 &  0.014 $\pm$  0.011 &  0.231 $\pm$  0.071 &  0.212 $\pm$  0.071  \\
HSK~67 & 08 19 26.70 & +70 41 55.40 & 10.0 & 0.100 $\pm$  0.007 &  0.110 $\pm$  0.006 &  0.144 $\pm$  0.011 &  0.177 $\pm$  0.009 &  0.090 $\pm$  0.009 &  2.570 $\pm$  0.156 &  2.534 $\pm$  0.156  \\
HSK~70 & 08 19 27.69 & +70 42 20.62 & 8.0 & 0.126 $\pm$  0.148 &  0.125 $\pm$  0.116 &  0.072 $\pm$  0.099 &  0.156 $\pm$  0.049 &  0.076 $\pm$  0.086 &  1.460 $\pm$  0.095 &  1.414 $\pm$  0.109  \\
HSK~71-73 & 08 19 28.98 & +70 43 01.00 & 10.0 & 0.253 $\pm$  0.012 &  0.386 $\pm$  0.013 &  0.846 $\pm$  0.021 &  1.370 $\pm$  0.025 &  0.932 $\pm$  0.025 & 25.500 $\pm$  0.378 & 25.408 $\pm$  0.378  \\
HSK~45 & 08 19 12.69 & +70 43 07.44 & 11.0 & .630 $\pm$  0.213 &  1.940 $\pm$  0.107 &  4.800 $\pm$  0.123 &  7.060 $\pm$  0.140 &  4.828 $\pm$  0.164 & 25.300 $\pm$  0.345 & 25.072 $\pm$  0.354  \\\hline
\multicolumn {10}{c}{IC~2574}   \\ \hline
IC~2574-1    & 10 28 55.571 & +68 27 54.7 & 8.0 & 0.019 $\pm$ 0.006 & 0.015
$\pm$ 0.006 & 0.063 $\pm$ 0.011 & 0.136 $\pm$ 0.007 & 0.105 $\pm$ 0.007 & 2.120 $\pm$ 0.045 & 2.117 $\pm$ 0.045\\
IC~2574-2    & 10 28 58.93 & +68 28 27.72 & 7.0 & 0.044 $\pm$ 0.004 & 0.045
$\pm$ 0.004 & 0.080 $\pm$ 0.007 & 0.151 $\pm$ 0.010 & 0.109 $\pm$ 0.009 & 1.410 $\pm$ 0.077 & 1.402 $\pm$ 0.077\\
IC~2574-3	& 10 28 48.35 & +68 28 03.14 & 9.0 & 0.301 $\pm$ 0.027 & 0.333
$\pm$ 0.018 & 0.970 $\pm$ 0.043 & 2.140 $\pm$ 0.079 & 1.648 $\pm$ 0.075 & 24.900 $\pm$ 0.529 & 24.850 $\pm$ 0.529\\
IC~2574-5	& 10 28 50.11 & +68 28 23.64 & 8.0 & 0.087 $\pm$ 0.020 & 0.080
$\pm$ 0.019 & 0.202 $\pm$ 0.024 & 0.384 $\pm$ 0.040 & 0.285 $\pm$ 0.038 & 3.020 $\pm$ 0.409 & 3.005 $\pm$ 0.409\\
IC~2574-6	& 10 28 48.82 & +68 28 34.96 & 7.0 & 0.054 $\pm$ 0.015 & 0.059
$\pm$ 0.015 & 0.087 $\pm$ 0.021 & 0.162 $\pm$ 0.035 & 0.113 $\pm$ 0.036 & 1.990 $\pm$ 0.271 & 1.980 $\pm$ 0.271\\
IC~2574-8	& 10 28 43.57 & +68 28 25.79 & 8.0 & 0.170 $\pm$ 0.022 & 0.215
$\pm$ 0.025 & 0.261 $\pm$ 0.025 & 0.405 $\pm$ 0.035 & 0.260 $\pm$ 0.033 & 6.480 $\pm$ 0.280 & 6.450 $\pm$ 0.280\\
IC~2574-9	& 10 28 37.25 & +68 28 00.71 & 8.0 & 0.089 $\pm$ 0.008 & 0.076
$\pm$ 0.007 & 0.110 $\pm$ 0.009 & 0.213 $\pm$ 0.009 & 0.149 $\pm$ 0.009 & 2.170 $\pm$ 0.076 & 2.154 $\pm$ 0.076\\
IC~2574-10	& 10 28 38.99 & +68 28 05.31 & 6.0 & 0.052 $\pm$ 0.011 & 0.047
$\pm$ 0.009 & 0.055 $\pm$ 0.011 & 0.050 $\pm$ 0.009 & 0.022 $\pm$ 0.009 & 0.239 $\pm$ 0.113 & 0.229 $\pm$ 0.113\\
IC~2574-11	& 10 28 39.97 & +68 28 31.01 & 8.0 & 0.008 $\pm$ 0.009 & 0.008
$\pm$ 0.008 & 0.011 $\pm$ 0.007 & 0.051 $\pm$ 0.008 & 0.039 $\pm$ 0.012 & 0.059 $\pm$ 0.093 & 0.058 $\pm$ 0.093\\
IC~2574-13	& 10 28 32.53 & +68 28 01.15 & 8.0 & 0.014 $\pm$ 0.009 & 0.006
$\pm$ 0.006 & 0.010 $\pm$ 0.007 & 0.008 $\pm$ 0.001 & 0.003 $\pm$ 0.001 & 0.116 $\pm$ 0.041 & 0.114 $\pm$ 0.041\\
IC~2574-14	& 10 28 30.71 & +68 28 08.27 & 7.0 & 0.031 $\pm$ 0.009 & 0.026
$\pm$ 0.007 & 0.053 $\pm$ 0.004 & 0.064 $\pm$ 0.005 & 0.040 $\pm$ 0.005 & 0.631 $\pm$ 0.033 & 0.625 $\pm$ 0.033 \\\hline
\multicolumn {10}{c}{NGC~628}   \\ \hline
NGC~628-7 & 1 36 47.257	& +15 45 50.21 & 10.0 &0.620 $\pm$  0.048 &  0.518 $\pm$  0.037 &  3.190 $\pm$  0.212 &  8.370 $\pm$  0.524 &  6.207 $\pm$  0.504 &  9.630 $\pm$  0.517 &  9.531 $\pm$  0.517  \\
NGC~628-8 & 1 36 51.220	& +15 45 58.80 & 6.0& 0.091 $\pm$  0.029 &  0.080 $\pm$  0.024 &  0.655 $\pm$  0.173 &  1.640 $\pm$  0.427 &  1.217 $\pm$  0.412 &  1.960 $\pm$  0.563 &  1.946 $\pm$  0.563  \\
NGC~628-10& 1 36 53.931	& +15 46 56.26 & 12.0 & 0.397 $\pm$  0.025 &  0.277 $\pm$  0.019 &  2.790 $\pm$  0.116 &  6.650 $\pm$  0.246 &  4.597 $\pm$  0.245 &  5.040 $\pm$  0.205 &  4.977 $\pm$  0.205  \\
+081-140& 1 36 47.257 &	+15 44 44.24 & 10.0 &0.209 $\pm$  0.025 &  0.164 $\pm$  0.017 &  0.670 $\pm$  0.128 &  1.590 $\pm$  0.313 &  1.159 $\pm$  0.299 &  2.080 $\pm$  0.242 &  2.047 $\pm$  0.242  \\
+062-158& 1 36 45.877 &	+15 44 24.55 & 11.0 & 0.118 $\pm$  0.030 &  0.101 $\pm$  0.020 &  0.901 $\pm$  0.165 &  2.460 $\pm$  0.448 &  1.711 $\pm$  0.445 &  1.630 $\pm$  0.328 &  1.611 $\pm$  0.328  \\
+047-172& 1 36 44.747 &	+15 44 05.46 & 14.0 & 0.155 $\pm$  0.037 &  0.128 $\pm$  0.026 &  1.140 $\pm$  0.180 &  3.230 $\pm$  0.443 &  2.371 $\pm$  0.428 &  2.870 $\pm$  0.269 &  2.845 $\pm$  0.269  \\
+044-175& 1 36 43.908 &	+15 43 49.04 & 12.0 &0.081 $\pm$  0.020 &  0.077 $\pm$  0.015 &  0.497 $\pm$  0.067 &  1.170 $\pm$  0.158 &  0.897 $\pm$  0.150 &  2.410 $\pm$  0.186 &  2.397 $\pm$  0.186  \\
+178-052& 1 36 53.945 &	+15 46 10.63 & 12.0& 0.145 $\pm$  0.014 &  0.120 $\pm$  0.010 &  1.450 $\pm$  0.083 &  2.690 $\pm$  0.119 &  1.823 $\pm$  0.121 &  2.630 $\pm$  0.161 &  2.607 $\pm$  0.161  \\
-086+186& 1 36 35.643 &	+15 50 07.64 & 11.0 &0.433 $\pm$  0.023 &  0.378 $\pm$  0.016 &  2.480 $\pm$  0.076 &  6.160 $\pm$  0.192 &  4.820 $\pm$  0.182 & 15.500 $\pm$  0.396 & 15.431 $\pm$  0.396  \\
-075+200& 1 36 36.803 &	+15 50 27.46 & 11.0& 0.305 $\pm$  0.020 &  0.233 $\pm$  0.017 &  1.700 $\pm$  0.109 &  4.600 $\pm$  0.300 &  3.473 $\pm$  0.292 &  5.750 $\pm$  0.627 &  5.701 $\pm$  0.627  \\
\hline
\multicolumn {10}{c}{NGC~925}   \\ \hline
+042-011 & 2 27 19.989 & +33 34 35.21 & 6.0 & 0.345 $\pm$  0.053 &  0.301 $\pm$  0.039 &  1.790 $\pm$  0.246 &  4.540 $\pm$  0.668 &  3.467 $\pm$  0.633 &  8.600 $\pm$  0.921 &  8.499 $\pm$  0.921  \\
+135-016 & 2 27 27.607 & +33 34 28.03 & 5.0 &0.083 $\pm$  0.019 &  0.063 $\pm$  0.013 &  0.313 $\pm$  0.049 &  0.740 $\pm$  0.100 &  0.547 $\pm$  0.095 &  1.210 $\pm$  0.129 &  1.186 $\pm$  0.129  \\
+087-031 & 2 27 22.709 & +33 34 15.91 & 5.0 &0.041 $\pm$  0.015 &  0.030 $\pm$  0.010 &  0.209 $\pm$  0.051 &  0.558 $\pm$  0.133 &  0.422 $\pm$  0.127 &  0.784 $\pm$  0.100 &  0.772 $\pm$  0.100  \\
-012-066 & 2 27 15.893 & +33 33 37.92 & 10.0&0.256 $\pm$  0.025 &  0.242 $\pm$  0.018 &  1.370 $\pm$  0.072 &  3.370 $\pm$  0.133 &  2.650 $\pm$  0.126 & 13.500 $\pm$  0.245 & 13.425 $\pm$  0.245  \\
-047-058 & 2 27 13.206 & +33 33 47.48 & 8.0 &0.082 $\pm$  0.008 &  0.065 $\pm$  0.005 &  0.384 $\pm$  0.018 &  0.927 $\pm$  0.058 &  0.674 $\pm$  0.056 &  1.120 $\pm$  0.067 &  1.096 $\pm$  0.067  \\
-137+056 & 2 27 08.015 & +33 35 46.95 & 7.0 &0.073 $\pm$  0.010 &  0.060 $\pm$  0.008 &  0.238 $\pm$  0.018 &  0.600 $\pm$  0.045 &  0.447 $\pm$  0.043 &  1.020 $\pm$  0.098 &  0.999 $\pm$  0.098  \\
-080+087 & 2 27 10.526 & +33 36 11.28 & 8.0 &0.232 $\pm$  0.017 &  0.205 $\pm$  0.013 &  1.180 $\pm$  0.053 &  2.900 $\pm$  0.133 &  2.257 $\pm$  0.125 &  8.690 $\pm$  0.224 &  8.622 $\pm$  0.224  \\
-114+087 & 2 27 07.676 & +33 36 13.08 & 6.0 &0.125 $\pm$  0.016 &  0.098 $\pm$  0.013 &  0.380 $\pm$  0.044 &  0.880 $\pm$  0.094 &  0.653 $\pm$  0.090 &  1.840 $\pm$  0.142 &  1.804 $\pm$  0.142  \\
-198-013 & 2 27 01.378 & +33 34 37.25 & 5.0&0.075 $\pm$  0.018 &  0.067 $\pm$  0.017 &  0.457 $\pm$  0.087 &  1.150 $\pm$  0.214 &  0.840 $\pm$  0.214 &  1.300 $\pm$  0.493 &  1.278 $\pm$  0.493  \\
-220+004 & 2 26 59.530 & +33 34 48.60 & 9.0 &0.298 $\pm$  0.018 &  0.314 $\pm$  0.019 &  1.460 $\pm$  0.053 &  3.290 $\pm$  0.131 &  2.540 $\pm$  0.123 & 13.500 $\pm$  0.322 & 13.413 $\pm$  0.322  \\
+156-114 & 2 27 29.301 & +33 35 03.86 &6.0 & 0.011 $\pm$  0.020 &  0.002 $\pm$  0.013 &  0.031 $\pm$  0.034 &  0.072 $\pm$  0.070 &  0.060 $\pm$  0.066 &  1.520 $\pm$  0.153 &  1.523 $\pm$  0.153  \\
+206-114 & 2 27 33.425 & +33 32 51.23 & 6.0& 0.014 $\pm$  0.004 &  0.017 $\pm$  0.004 &  0.043 $\pm$  0.008 &  0.097 $\pm$  0.013 &  0.071 $\pm$  0.012 &  0.276 $\pm$  0.027 &  0.272 $\pm$  0.027  \\
+019+143 & 2 27 18.448 & +33 37 08.92 & 7.0 &0.012 $\pm$  0.005 &  0.011 $\pm$  0.004 &  0.056 $\pm$  0.009 &  0.098 $\pm$  0.008 &  0.073 $\pm$  0.008 &  0.500 $\pm$  0.035 &  0.496 $\pm$  0.035  \\
-250+019 & 2 26 56.807 & +33 35 03.72 & 7.0 &0.092 $\pm$  0.007 &  0.081 $\pm$  0.006 &  0.456 $\pm$  0.020 &  1.130 $\pm$  0.048 &  0.889 $\pm$  0.045 &  4.570 $\pm$  0.108 &  4.543 $\pm$  0.108  \\
-274+010 & 2 26 54.961 & +33 34 52.44 & 5.0 &0.056 $\pm$  0.022 &  0.054 $\pm$  0.018 &  0.105 $\pm$  0.028 &  0.209 $\pm$  0.044 &  0.147 $\pm$  0.042 &  0.884 $\pm$  0.104 &  0.868 $\pm$  0.104  \\
-174+140 & 2 27 03.010 & +33 37 04.31 & 5.0 &0.226 $\pm$  0.099 &  0.140 $\pm$  0.059 &  0.223 $\pm$  0.051 &  0.361 $\pm$  0.036 &  0.204 $\pm$  0.057 &  0.531 $\pm$  0.049 &  0.465 $\pm$  0.057  \\
-159+162 & 2 27 04.215 & +33 37 27.09 & 5.0&0.005 $\pm$  0.001 &  0.004 $\pm$  0.001 & 0.001 $\pm$  0.007 &  0.041 $\pm$  0.011 &    0.025 $\pm$    0.011 &  0.038 $\pm$  0.028 &  0.037 $\pm$  0.028  \\
-149+177 & 2 27 04.848 & +33 37 42.12 & 7.0 &0.030 $\pm$  0.003 &  0.026 $\pm$  0.003 &  0.129 $\pm$  0.008 &  0.284 $\pm$  0.020 &  0.209 $\pm$  0.019 &  0.510 $\pm$  0.049 &  0.501 $\pm$  0.049  \\
-262+011 & 2 26 55.744 & +33 34 55.17 & 5.0 &0.032 $\pm$  0.018 &  0.029 $\pm$  0.016 &  0.045 $\pm$  0.025 &  0.069 $\pm$  0.048 &  0.032 $\pm$  0.070 &  0.048 $\pm$  0.128 &  0.038 $\pm$  0.128  \\
\hline
\multicolumn {10}{c}{NGC~3184}   \\ \hline
-058-007 & 10 18 12.939 & +41 25 19.42 & 5.0 & 0.003 $\pm$  0.008 &  0.005 $\pm$  0.006 &  0.143 $\pm$  0.038 &  0.426 $\pm$  0.116 &  0.327 $\pm$  0.111 &  0.477 $\pm$  0.102 &  0.476 $\pm$  0.102  \\
-064-006 & 10 18 11.652 & +41 25 19.42 & 5.0 & 0.072 $\pm$  0.014 &  0.056 $\pm$  0.011 &  0.453 $\pm$  0.072 &  1.190 $\pm$  0.209 &  0.900 $\pm$  0.198 &  1.750 $\pm$  0.190 &  1.723 $\pm$  0.190  \\
-080-005 & 10 18 09.884 & +41 25 21.22 & 5.0 & 0.070 $\pm$  0.017 &  0.054 $\pm$  0.013 &  0.481 $\pm$  0.123 &  1.260 $\pm$  0.345 &   0.935 $\pm$  0.331 &  1.490 $\pm$  0.436 &  1.463 $\pm$  0.436  \\
+085-004 & 10 18 23.794 & +41 26 00.87 & 7.0 & 0.084 $\pm$  0.024 &  0.065 $\pm$  0.017 &  0.749 $\pm$  0.118 &  1.940 $\pm$  0.346 &  1.450 $\pm$  0.330 &  2.380 $\pm$  0.263 &  2.348 $\pm$  0.263  \\
+079+035 & 10 18 23.795 & +41 26 32.12 & 7.0 & 0.104 $\pm$  0.033 &  0.076 $\pm$  0.021 &  0.766 $\pm$  0.073 &  2.090 $\pm$  0.180 &  1.457 $\pm$  0.179 &  1.420 $\pm$  0.130 &  1.380 $\pm$  0.131  \\
+074+064 & 10 18 23.425 & +41 26 56.08 & 12.0 & 0.463 $\pm$  0.067 &  0.434 $\pm$  0.046 &  2.720 $\pm$  0.229 &  6.870 $\pm$  0.644 &  5.434 $\pm$  0.605 & 33.700 $\pm$  0.571 & 33.523 $\pm$  0.572  \\
+059-079 & 10 18 22.229 & +41 24 10.52 & 7.0 & 0.084 $\pm$  0.023 &  0.069 $\pm$  0.016 &  0.421 $\pm$  0.092 &  1.100 $\pm$  0.272 &  0.792 $\pm$  0.260 &  1.090 $\pm$  0.148 &  1.058 $\pm$  0.148  \\
+092-093 & 10 18 25.090 & +41 23 55.02 & 6.0 & 0.092 $\pm$  0.011 &  0.074 $\pm$  0.008 &  0.513 $\pm$  0.036 &  1.400 $\pm$  0.097 &  0.965 $\pm$  0.096 &  0.936 $\pm$  0.081 &  0.901 $\pm$  0.081  \\
+111-102 & 10 18 26.540 & +41 23 43.98 & 8.0 & 0.018 $\pm$  0.011 &  0.009 $\pm$  0.008 &  0.028 $\pm$  0.030 &  0.073 $\pm$  0.067 &  0.047 $\pm$  0.067 &  0.038 $\pm$  0.048 &  0.031 $\pm$  0.048  \\
+005+135 & 10 18 17.283 & +41 27 42.55 & 8.0 & 0.223 $\pm$  0.020 &  0.170 $\pm$  0.013 &  1.320 $\pm$  0.069 &  3.200 $\pm$  0.179 &  2.369 $\pm$  0.171 &  4.470 $\pm$  0.147 &  4.385 $\pm$  0.147  \\
-017+137 & 10 18 15.430 & +41 27 43.59 & 8.0 & 0.116 $\pm$  0.016 &  0.098 $\pm$  0.010 &  0.749 $\pm$  0.046 &  1.890 $\pm$  0.105 &  1.348 $\pm$  0.102 &  1.760 $\pm$  0.076 &  1.716 $\pm$  0.076  \\
\hline
\multicolumn {10}{c}{NGC~3621}   \\ \hline
s5a2 & 11 18 20.861	& -32 48 51.97 & 7.0 & 0.162 $\pm$  0.087 &  0.155 $\pm$  0.061 &  1.300 $\pm$  0.293 &  3.540 $\pm$  0.755 &  2.864 $\pm$  0.710 & 14.200 $\pm$  0.743 & 14.173 $\pm$  0.743  \\
s5a1 & 11 18 18.838	& -32 47 42.59 & 9.0 & 1.280 $\pm$  0.113 &  1.260 $\pm$  0.078 &  7.660 $\pm$  0.370 & 18.700 $\pm$  0.881 & 14.855 $\pm$  0.829 & 76.900 $\pm$  1.190 & 76.689 $\pm$  1.190  \\
p1a8 & 11 18 13.589	& -32 47 28.51 & 9.0 & 0.781 $\pm$  0.180 &  0.703 $\pm$  0.132 &  5.660 $\pm$  0.749 & 14.900 $\pm$  1.910 & 11.572 $\pm$  1.809 & 27.200 $\pm$  1.900 & 27.071 $\pm$  1.900  \\
p2a6 & 11 18 14.707	& -32 48 24.56 & 6.0 & 0.163 $\pm$  0.194 &  0.131 $\pm$  0.125 &  0.880 $\pm$  0.385 &  2.480 $\pm$  0.964 &  1.857 $\pm$  0.926 &  2.720 $\pm$  0.736 &  2.693 $\pm$  0.737  \\
p1a9 & 11 18 08.489	& -32 47 43.11 & 5.0 & 0.037 $\pm$  0.036 &  0.029 $\pm$  0.024 &  0.004 $\pm$  0.049 &  0.050 $\pm$  0.092 &  0.037 $\pm$  0.089 &  0.200 $\pm$  0.065 &  0.194 $\pm$  0.065  \\
p2a7 & 11 18 10.550	& -32 48 00.63 & 7.0 & 0.027 $\pm$  0.036 &  0.023 $\pm$  0.025 &  1.420 $\pm$  0.094 &  3.590 $\pm$  0.251 &  2.774 $\pm$  0.239 &  5.510 $\pm$  0.224 &  5.506 $\pm$  0.224  \\
p1a7 & 11 18 12.094	& -32 48 23.85 & 7.0 & 0.064 $\pm$  0.098 &  0.074 $\pm$  0.067 &  0.898 $\pm$  0.314 &  2.500 $\pm$  0.813 &  1.891 $\pm$  0.783 &  2.880 $\pm$  0.960 &  2.869 $\pm$  0.960  \\
p1a6 & 11 18 18.638	& -32 48 15.25 & 7.0 & 0.742 $\pm$  0.198 &  0.653 $\pm$  0.132 &  5.530 $\pm$  0.533 & 14.000 $\pm$  1.330 & 10.403 $\pm$  1.274 & 16.600 $\pm$  0.999 & 16.478 $\pm$  1.000  \\
p2a5 & 11 18 13.507	& -32 49 08.33 & 5.0 & 0.118 $\pm$  0.082 &  0.107 $\pm$  0.055 &  0.799 $\pm$  0.143 &  2.170 $\pm$  0.327 &  1.572 $\pm$  0.319 &  1.910 $\pm$  0.226 &  1.891 $\pm$  0.226  \\
p1a5 & 11 18 18.178	& -32 48 55.75 & 7.0 & 1.090 $\pm$  0.309 &  0.872 $\pm$  0.200 &  5.760 $\pm$  0.856 & 15.600 $\pm$  2.380 & 11.691 $\pm$  2.284 & 18.500 $\pm$  2.690 & 18.320 $\pm$  2.690  \\
p2a4 & 11 18 18.929	& -32 49 10.57 & 7.0 & 0.877 $\pm$  0.245 &  0.711 $\pm$  0.163 &  5.310 $\pm$  1.020 & 14.000 $\pm$  2.740 & 10.691 $\pm$  2.604 & 20.800 $\pm$  2.570 & 20.655 $\pm$  2.570  \\
p1a3 & 11 18 18.830	& -32 50 16.04 & 8.0 & 0.342 $\pm$  0.085 &  0.255 $\pm$  0.058 &  1.820 $\pm$  0.274 &  4.670 $\pm$  0.718 &  3.488 $\pm$  0.685 &  5.790 $\pm$  0.430 &  5.734 $\pm$  0.430  \\
s1a4 & 11 18 22.207	& -32 51 40.22 & 5.0 & 0.024 $\pm$  0.071 &  0.025 $\pm$  0.056 &  0.331 $\pm$  0.267 &  0.963 $\pm$  0.195 &  0.754 $\pm$  0.202 &  1.500 $\pm$  0.173 &  1.496 $\pm$  0.173  \\
p2a2 & 11 18 21.480	& -32 50 05.82 & 5.0 & 0.127 $\pm$  0.027 &  0.095 $\pm$  0.019 &  0.958 $\pm$  0.135 &  2.470 $\pm$  0.358 &  1.912 $\pm$  0.339 &  4.290 $\pm$  0.416 &  4.269 $\pm$  0.416  \\
s2a2 & 11 18 21.842	& -32 49 54.36 & 6.0 & 0.159 $\pm$  0.046 &  0.152 $\pm$  0.032 &  0.804 $\pm$  0.177 &  2.030 $\pm$  0.449 &  1.557 $\pm$  0.426 &  3.840 $\pm$  0.609 &  3.814 $\pm$  0.609  \\
p2a1 & 11 18 25.442	& -32 50 26.44 & 5.0 & 0.021 $\pm$  0.007 &  0.017 $\pm$  0.004 &  0.154 $\pm$  0.022 &  0.318 $\pm$  0.042 &  0.236 $\pm$  0.041 &  0.538 $\pm$  0.070 &  0.535 $\pm$  0.070  \\
p1a2 & 11 18 23.815	& -32 50 04.60 & 5.0 & 0.134 $\pm$  0.014 &  0.113 $\pm$  0.010 &  0.442 $\pm$  0.049 &  1.010 $\pm$  0.123 &  0.752 $\pm$  0.117 &  1.850 $\pm$  0.133 &  1.828 $\pm$  0.133  \\
s2a3 & 11 18 18.765	& -32 49 24.59 & 6.0 & 0.515 $\pm$  0.222 &  0.398 $\pm$  0.171 &  2.160 $\pm$  1.070 &  5.750 $\pm$  2.830 &  4.312 $\pm$  2.732 &  7.360 $\pm$  4.840 &  7.275 $\pm$  4.840  \\
p1a4 & 11 18 21.515	& -32 49 09.27 & 7.0 & 0.661 $\pm$  0.124 &  0.584 $\pm$  0.093 &  3.560 $\pm$  0.295 &  8.960 $\pm$  0.769 &  7.097 $\pm$  0.725 & 29.800 $\pm$  1.460 & 29.691 $\pm$  1.460  \\
s3a2 & 11 18 30.554	& -32 48 47.33 & 6.0 & 0.034 $\pm$  0.004 &  0.021 $\pm$  0.003 &  0.027 $\pm$  0.012 &  0.067 $\pm$  0.022 &  0.046 $\pm$  0.021 &  0.087 $\pm$  0.050 &  0.082 $\pm$  0.050  \\
s3a1 & 11 18 30.425	& -32 48 34.15 & 5.0 & 0.013 $\pm$  0.004 &  0.011 $\pm$  0.003 &  0.081 $\pm$  0.012 &  0.159 $\pm$  0.029 &  0.122 $\pm$  0.027 &  0.561 $\pm$  0.068 &  0.559 $\pm$  0.069  \\
\hline
\multicolumn {10}{c}{NGC~6946}   \\ \hline
NGC~6946-2 & 20 34 34.935	& +60 11 38.62 & 6.0& 0.965 $\pm$  0.143 &  0.833 $\pm$  0.113 &  6.040 $\pm$  0.791 & 15.000 $\pm$  1.960 & 11.659 $\pm$  1.856 & 32.100 $\pm$  3.290 & 31.959 $\pm$  3.290  \\
NGC~6946-6 & 20 34 52.435	& +60 10 18.56 & 6.0& 0.131 $\pm$  0.140 &  0.078 $\pm$  0.088 &  0.539 $\pm$  0.305 &  1.640 $\pm$  0.898 &  0.941 $\pm$  1.309 &  0.449 $\pm$  0.758 &  0.430 $\pm$  0.758  \\
NGC~6946-8 & 20 34 54.399	& +60 10 38.86 & 6.0& 0.542 $\pm$  0.090 &  0.489 $\pm$  0.073 &  2.680 $\pm$  0.688 &  6.720 $\pm$  1.920 &  4.985 $\pm$  1.832 &  8.610 $\pm$  1.730 &  8.531 $\pm$  1.730  \\
NGC~6946-10 & 20 34 58.778 & 	+60 10 50.60 & 5.0& 0.428 $\pm$  0.099 &  0.349 $\pm$  0.071 &  3.080 $\pm$  0.446 &  8.360 $\pm$  1.200 &  6.332 $\pm$  1.153 & 10.400 $\pm$  1.620 & 10.337 $\pm$  1.620  \\
NGC~6946-11 & 20 35 09.377 & 	+60 11 11.30 & 5.0& 0.091 $\pm$  0.040 &  0.072 $\pm$  0.031 &  0.578 $\pm$  0.245 &  1.560 $\pm$  0.604 &  1.183 $\pm$  0.577 &  2.010 $\pm$  0.570 &  1.997 $\pm$  0.570  \\
NGC~6946-21 & 20 34 36.086 & 	+60 11 41.99 & 6.0& 0.795 $\pm$  0.091 &  0.672 $\pm$  0.077 &  4.520 $\pm$  0.544 & 11.200 $\pm$  1.360 &  8.628 $\pm$  1.294 & 21.500 $\pm$  2.770 & 21.384 $\pm$  2.770  \\
NGC~6946-23 & 20 34 55.894 & 	+60 10 56.50 & 5.0& 0.278 $\pm$  0.122 &  0.253 $\pm$  0.090 &  1.780 $\pm$  0.577 &  4.460 $\pm$  1.510 &  3.446 $\pm$  1.429 &  8.600 $\pm$  1.480 &  8.559 $\pm$  1.480  \\
NGC~6946-26 & 20 35 07.970 & 	+60 11 13.01 & 5.0& 0.259 $\pm$  0.091 &  0.231 $\pm$  0.074 &  1.540 $\pm$  0.359 &  3.840 $\pm$  0.891 &  2.842 $\pm$  0.871 &  4.670 $\pm$  1.640 &  4.632 $\pm$  1.640  \\
NGC~6946-27 & 20 35 06.596 & 	+60 11 11.50 & 6.0& 0.335 $\pm$  0.136 &  0.339 $\pm$  0.111 &  2.760 $\pm$  0.829 &  7.340 $\pm$  2.100 &  5.613 $\pm$  2.015 & 10.700 $\pm$  3.690 & 10.651 $\pm$  3.690  \\
NGC~6946-30 & 20 35 16.805 & 	+60 11 00.39 & 9.0& 0.959 $\pm$  0.748 &  0.834 $\pm$  0.592 &  8.690 $\pm$  1.440 & 22.800 $\pm$  1.890 & 18.171 $\pm$  1.819 & 61.500 $\pm$  1.370 & 61.360 $\pm$  1.374  \\
NGC~6946-32 & 20 35 19.315 & 	+60 10 46.27 & 4.0& 0.066 $\pm$  0.019 &  0.053 $\pm$  0.014 &  0.298 $\pm$  0.070 &  0.720 $\pm$  0.179 &  0.502 $\pm$  0.176 &  0.609 $\pm$  0.154 &  0.599 $\pm$  0.154  \\
NGC~6946-34 & 20 34 45.728 & 	+60 08 20.58 & 4.0& 0.123 $\pm$  0.047 &  0.097 $\pm$  0.038 &  0.722 $\pm$  0.263 &  1.900 $\pm$  0.662 &  1.481 $\pm$  0.630 &  3.670 $\pm$  1.410 &  3.652 $\pm$  1.410  \\
NGC~6946-36 & 20 34 54.574 & 	+60 07 39.80 & 7.0& 0.863 $\pm$  0.149 &  0.719 $\pm$  0.108 &  5.450 $\pm$  0.554 & 13.900 $\pm$  1.420 & 10.798 $\pm$  1.344 & 27.400 $\pm$  1.510 & 27.274 $\pm$  1.510  \\
NGC~6946-39 & 20 35 05.107 & 	+60 07 03.53 & 7.0& 0.100 $\pm$  0.023 &  0.071 $\pm$  0.017 &  0.649 $\pm$  0.117 &  1.810 $\pm$  0.328 &  1.245 $\pm$  0.325 &  1.080 $\pm$  0.184 &  1.065 $\pm$  0.184  \\
\hline
\label{table:photometry}
\end{longtable}
\end{landscape}

\begin{longtable}{@{ } l @{ }lll}
\caption{Parameters of H~{\sc ii} complexes: designation, age, metallicity estimated by `ONS' method (except for Holmberg~II where the `NS' method is used) and the ratio [O~{\sc iii}]$\lambda 5007/H\beta$ considered as a potential indicator of the radiation field hardness}. \\ \hline
Object & age, Myr & 12+log(O/H) & [O~{\sc iii}]$\lambda 5007/H\beta$ \\
\hline \endhead

\multicolumn {4}{c}{IC~1727}   \\ \hline
IC~1727\_3  & $5.16 \pm 0.25$ & $8.01 \pm 0.02$& $1.77 \pm 0.26$\\
IC~1727\_5  & $4.99 \pm 0.24$ & $8.03 \pm 0.03$& $1.50 \pm 0.06$\\
IC~1727\_6  & $4.88 \pm 0.24$ & $8.02 \pm 0.06$& $1.48 \pm 0.09$\\
IC~1727\_9  & $6.85 \pm 0.27$ & $7.93 \pm 0.10$& $1.26 \pm 0.21$\\
IC~1727\_10 & $3.89 \pm 0.18$ & $8.05 \pm 0.03$& $2.32 \pm 0.24$\\
IC~1727\_11 & $3.79 \pm 0.17$ & $8.11 \pm 0.14$& $3.80 \pm 0.19$\\ \hline
\multicolumn {4}{c}{NGC~7741}   \\ \hline
NGC~7741\_1 & $5.18 \pm 0.15$ & $8.13 \pm 0.06$& $1.78 \pm 0.36$\\
NGC~7741\_3 & $6.56 \pm 0.31$ & $8.49 \pm 0.01$& $0.72 \pm 0.48$\\
NGC~7741\_4 & $7.55 \pm 0.37$ & $8.37 \pm 0.03$& $0.85 \pm 0.08$\\
NGC~7741\_7 & $6.58 \pm 0.31$ & $8.28 \pm 0.11$& $0.87 \pm 0.08$\\
NGC~7741\_9 & $8.55 \pm 0.25$ & $8.35 \pm 0.05$& $1.76 \pm 0.12$\\
NGC~7741\_10& $6.55 \pm 0.42$ & $8.32 \pm 0.06$& $2.83 \pm 0.36$\\ \hline
\multicolumn {4}{c}{Holmberg II}   \\ \hline
HSK~7      & $3.69 \pm 0.14$ & $7.76 \pm 0.03$& $4.25 \pm 0.18$\\
HSK~15      & $5.21 \pm 0.14$ & $7.63 \pm 0.10$& $1.60 \pm 0.08$\\
HSK~16-17   & $5.59 \pm 0.15$ & $7.68 \pm 0.05$& $2.06 \pm 0.07$\\
HSK~20     & $5.73 \pm 0.16$ & $7.45 \pm 0.04$& $1.01 \pm 0.10$\\
HSK~25      & $6.64 \pm 0.18$ & $7.53 \pm 0.02$& $1.13 \pm 0.02$\\
HSK~31      & $6.38 \pm 0.18$ & $7.97 \pm 0.12$& $3.29 \pm 0.04$\\
HSK~59      & $3.83 \pm 0.12$ & $7.60 \pm 0.03$& $1.82 \pm 0.02$\\
HSK~61      & $7.81 \pm 0.22$ & $7.66 \pm 0.03$& $2.06 \pm 0.03$\\
HSK~67      & $3.74 \pm 0.11$ & $7.57 \pm 0.02$& $2.97 \pm 0.02$\\
HSK~70      & $4.82 \pm 0.14$ & $7.77 \pm 0.02$& $2.93 \pm 0.02$\\
HSK~71-73   & $3.58 \pm 0.10$ & $7.73 \pm 0.03$& $2.23 \pm 0.02$\\
\hline
\multicolumn {4}{c}{IC~2574}   \\ \hline
IC~2574-1   & $3.41 \pm 0.10$ & $7.72 \pm 0.03$& $4.41 \pm 0.33$\\
IC~2574-2   & $4.21 \pm 0.13$ & $7.78 \pm 0.03$& $2.69 \pm 0.20$\\
IC~2574-3   & $2.71 \pm 0.08$ & $7.82 \pm 0.03$& $4.69 \pm 0.35$\\
IC~2574-5   & $4.03 \pm 0.12$ & $7.81 \pm 0.04$& $4.57 \pm 0.34$\\
IC~2574-6   & $3.97 \pm 0.12$ & $7.77 \pm 0.03$& $4.12 \pm 0.31$\\
IC~2574-8   & $3.93 \pm 0.12$ & $7.80 \pm 0.03$& $4.15 \pm 0.31$\\
IC~2574-9   & $4.66 \pm 0.14$ & $7.85 \pm 0.06$& $3.46 \pm 0.26$\\
IC~2574-10  & $5.06 \pm 0.14$ & $7.72 \pm 0.04$& $3.70 \pm 0.28$\\
IC~2574-11  & $4.53 \pm 0.14$ & $7.80 \pm 0.15$& $2.64 \pm 0.20$\\
IC~2574-13  & $4.36 \pm 0.14$ & $7.76 \pm 0.15$& $3.71 \pm 0.28$\\
IC~2574-14  & $5.82 \pm 0.15$ & $7.79 \pm 0.15$& $3.92 \pm 0.30$\\ \hline
\multicolumn {4}{c}{NGC~628}   \\ \hline
NGC~628-7    &$ 3.37 \pm 0.23 $ & $ 8.41 \pm 0.04 $ & $ 0.49 \pm 0.07 $\\
NGC~628-8    &$ 4.43 \pm 0.29 $ & $ 8.45 \pm 0.02 $ & $ 0.66 \pm 0.07 $\\
NGC~628-10   &$ 4.88 \pm 0.35 $ & $ 8.37 \pm 0.12 $ & $ 0.36 \pm 0.10 $\\
+081-140 &$ 4.61 \pm 0.31 $ & $ 8.46 \pm 0.03 $ & $ 1.11 \pm 0.03 $\\
+062-158 &$ 4.55 \pm 0.32 $ & $ 8.60 \pm 0.03 $ & $ 0.26 \pm 0.04 $\\
+047-172 &$ 5.00 \pm 0.34 $ & $ 8.51 \pm 0.01 $ & $ 0.77 \pm 0.03 $\\
+044-175 &$ 3.24 \pm 0.22 $ & $ 8.48 \pm 0.01 $ & $ 1.54 \pm 0.05 $\\
+178-052 &$ 3.88 \pm 0.26 $ & $ 8.46 \pm 0.04 $ & $ 0.85 \pm 0.03 $\\
-086+186 &$ 3.21 \pm 0.22 $ & $ 8.50 \pm 0.01 $ & $ 1.54 \pm 0.04 $\\
-075+200 &$ 4.86 \pm 0.33 $ & $ 8.48 \pm 0.01 $ & $ 1.10 \pm 0.05 $\\
\hline
\multicolumn {4}{c}{NGC~925}   \\ \hline
+042-011 &$ 5.01 \pm 0.33 $ & $ 8.42 \pm 0.01 $ & $ 0.95 \pm 0.03 $\\
+135-016 &$ 5.06 \pm 0.21 $ & $ 8.16 \pm 0.02 $ & $ 2.02 \pm 0.06 $\\
+087-031 &$ 4.41 \pm 0.28 $ & $ 8.36 \pm 0.01 $ & $ 1.70 \pm 0.05 $\\
-012-066 &$ 2.57 \pm 0.11 $ & $ 8.27 \pm 0.01 $ & $ 3.82 \pm 0.12 $\\
-047-058 &$ 4.79 \pm 0.31 $ & $ 8.32 \pm 0.01 $ & $ 1.15 \pm 0.04 $\\
-137+056 &$ 3.03 \pm 0.20 $ & $ 8.40 \pm 0.01 $ & $ 1.79 \pm 0.05 $\\
-080+087 &$ 3.94 \pm 0.26 $ & $ 8.32 \pm 0.01 $ & $ 1.45 \pm 0.04 $\\
-114+087 &$ 5.94 \pm 0.25 $ & $ 8.21 \pm 0.02 $ & $ 1.98 \pm 0.06 $\\
-198-013 &$ 3.77 \pm 0.15 $ & $ 8.13 \pm 0.02 $ & $ 3.41 \pm 0.10 $\\
-220+004 &$ 3.67 \pm 0.17 $ & $ 8.28 \pm 0.01 $ & $ 1.49 \pm 0.04 $\\
+156-114 &$ 4.07 \pm 0.17 $ & $ 8.27 \pm 0.01 $ & $ 1.65 \pm 0.05 $\\
+206-114 &$ 4.31 \pm 0.17 $ & $ 8.09 \pm 0.01 $ & $ 2.22 \pm 0.07 $\\
+019+143 &$ 3.80 \pm 0.15 $ & $ 8.07 \pm 0.02 $ & $ 4.43 \pm 0.14 $\\
-250+019 &$ 3.35 \pm 0.14 $ & $ 8.21 \pm 0.01 $ & $ 2.92 \pm 0.09 $\\
-274+010 &$ 5.03 \pm 0.21 $ & $ 8.17 \pm 0.02 $ & $ 2.87 \pm 0.09 $\\
-174+140 &$ 5.28 \pm 0.22 $ & $ 8.10 \pm 0.02 $ & $ 2.74 \pm 0.09 $\\
-159+162 &$ 5.61 \pm 0.22 $ & $ 8.03 \pm 0.02 $ & $ 0.99 \pm 0.05 $\\
-149+177 &$ 4.81 \pm 0.27 $ & $ 7.99 \pm 0.02 $ & $ 1.40 \pm 0.04 $\\
-262+011 &$ 5.12 \pm 0.21 $ & $ 8.11 \pm 0.03 $ & $ 1.97 \pm 0.09 $\\
\hline
\multicolumn {4}{c}{NGC~3184}   \\ \hline
-058-007 &$ 5.67 \pm 0.90 $ & $ 8.78 \pm 0.01 $ & $ 0.08 \pm 0.01 $\\
-064-006 &$ 5.36 \pm 0.85 $ & $ 8.70 \pm 0.01 $ & $ 0.14 \pm 0.02 $\\
-080-005 &$ 4.96 \pm 0.34 $ & $ 8.68 \pm 0.01 $ & $ 0.17 \pm 0.01 $\\
+085-004 &$ 0.10 \pm 0.01 $ & $ 8.65 \pm 0.01 $ & $ 0.17 \pm 0.02 $\\
+079+035 &$ 4.97 \pm 0.79 $ & $ 8.72 \pm 0.01 $ & $ 0.10 \pm 0.01 $\\
+074+064 &$ 6.80 \pm 0.46 $ & $ 8.61 \pm 0.01 $ & $ 0.29 \pm 0.03 $\\
+059-079 &$ 5.12 \pm 0.35 $ & $ 8.65 \pm 0.01 $ & $ 0.14 \pm 0.01 $\\
+092-093 &$ 4.52 \pm 0.32 $ & $ 8.64 \pm 0.01 $ & $ 0.15 \pm 0.01 $\\
+111-102 &$ 6.78 \pm 0.46 $ & $ 8.43 \pm 0.02 $ & $ 0.81 \pm 0.10 $\\
+005+135 &$ 4.71 \pm 0.32 $ & $ 8.48 \pm 0.01 $ & $ 0.78 \pm 0.02 $\\
-017+137 &$ 3.48 \pm 0.23 $ & $ 8.59 \pm 0.01 $ & $ 0.26 \pm 0.01 $\\
\hline
\multicolumn {4}{c}{NGC~3621}   \\ \hline
s5a2 &$ 4.94 \pm 0.33 $ & $ 8.45 \pm 0.03 $ & $ 1.45 \pm 0.20 $\\
s5a1 &$ 4.38 \pm 0.30 $ & $ 8.52 \pm 0.02 $ & $ 0.60 \pm 0.08 $\\
p1a8 &$ 6.20 \pm 0.40 $ & $ 8.51 \pm 0.05 $ & $ 0.65 \pm 0.23 $\\
p2a6 &$ 4.76 \pm 0.33 $ & $ 8.63 \pm 0.05 $ & $ 0.21 \pm 0.07 $\\
p1a9 &$ 3.14 \pm 0.21 $ & $ 8.37 \pm 0.09 $ & $ 2.13 \pm 0.22 $\\
p2a7 &$ 4.68 \pm 0.31 $ & $ 8.47 \pm 0.02 $ & $ 1.17 \pm 0.12 $\\
p1a7 &$ 3.14 \pm 0.21 $ & $ 8.47 \pm 0.02 $ & $ 0.66 \pm 0.07 $\\
p1a6 &$ 2.19 \pm 0.16 $ & $ 8.53 \pm 0.03 $ & $ 0.37 \pm 0.11 $\\
p2a5 &$ 4.36 \pm 0.32 $ & $ 8.69 \pm 0.06 $ & $ 0.10 \pm 0.05 $\\
p1a5 &$ 5.10 \pm 0.84 $ & $ 8.91 \pm 0.05 $ & $ 0.02 \pm 0.01 $\\
p2a4 &$ 5.60 \pm 0.37 $ & $ 8.60 \pm 0.03 $ & $ 0.20 \pm 0.05 $\\
p1a3 &$ 3.02 \pm 0.21 $ & $ 8.51 \pm 0.04 $ & $ 1.49 \pm 0.22 $\\
s1a4 &$ 3.35 \pm 0.23 $ & $ 8.38 \pm 0.02 $ & $ 2.53 \pm 0.25 $\\
p2a2 &$ 4.67 \pm 0.31 $ & $ 8.45 \pm 0.02 $ & $ 0.93 \pm 0.10 $\\
s2a2 &$ 4.86 \pm 0.33 $ & $ 8.46 \pm 0.02 $ & $ 0.69 \pm 0.07 $\\
p2a1 &$ 4.21 \pm 0.28 $ & $ 8.46 \pm 0.01 $ & $ 0.95 \pm 0.17 $\\
p1a2 &$ 5.48 \pm 0.37 $ & $ 8.40 \pm 0.02 $ & $ 2.01 \pm 0.20 $\\
s2a3 &$ 5.29 \pm 0.35 $ & $ 8.60 \pm 0.02 $ & $ 0.19 \pm 0.03 $\\
p1a4 &$ 4.53 \pm 0.30 $ & $ 8.47 \pm 0.02 $ & $ 1.15 \pm 0.11 $\\
s3a2 &$ 5.20 \pm 0.35 $ & $ 8.49 \pm 0.02 $ & $ 2.59 \pm 0.26 $\\
s3a1 &$ 4.81 \pm 0.32 $ & $ 8.41 \pm 0.01 $ & $ 2.29 \pm 0.23 $\\
\hline
\multicolumn {4}{c}{NGC~6946}   \\ \hline
NGC~6946-2 &$ 4.47 \pm 0.30 $ & $ 8.47 \pm 0.01 $ & $ 0.79 \pm 0.03 $\\
NGC~6946-6 &$ 7.26 \pm 0.49 $ & $ 8.59 \pm 0.02 $ & $ 0.35 \pm 0.04 $\\
NGC~6946-8 &$ 4.75 \pm 0.33 $ & $ 8.60 \pm 0.01 $ & $ 0.29 \pm 0.01 $\\
NGC~6946-10 &$ 6.35 \pm 1.03 $ & $ 8.76 \pm 0.05 $ & $ 0.04 \pm 0.02 $\\
NGC~6946-11 &$ 4.41 \pm 0.32 $ & $ 8.69 \pm 0.02 $ & $ 0.07 \pm 0.02 $\\
NGC~6946-21 &$ 3.93 \pm 0.26 $ & $ 8.41 \pm 0.01 $ & $ 0.31 \pm 0.02 $\\
NGC~6946-23 &$ 4.60 \pm 0.31 $ & $ 8.47 \pm 0.01 $ & $ 0.63 \pm 0.04 $\\
NGC~6946-26 &$ 5.37 \pm 0.36 $ & $ 8.37 \pm 0.01 $ & $ 0.74 \pm 0.03 $\\
NGC~6946-27 &$ 4.37 \pm 0.28 $ & $ 8.35 \pm 0.01 $ & $ 0.95 \pm 0.03 $\\
NGC~6946-30 &$ 4.55 \pm 0.29 $ & $ 8.38 \pm 0.01 $ & $ 2.02 \pm 0.05 $\\
NGC~6946-32 &$ 6.57 \pm 0.44 $ & $ 8.40 \pm 0.02 $ & $ 1.56 \pm 0.11 $\\
NGC~6946-34 &$ 5.76 \pm 0.38 $ & $ 8.67 \pm 0.02 $ & $ 0.19 \pm 0.03 $\\
NGC~6946-36 &$ 6.39 \pm 0.43 $ & $ 8.32 \pm 0.01 $ & $ 0.08 \pm 0.02 $\\
NGC~6946-39 &$ 4.57 \pm 0.33 $ & $ 8.65 \pm 0.01 $ & $ 0.15 \pm 0.01 $\\
\hline

\label{table: key results}
\end{longtable}

\end{document}